\documentclass[]{aastex701}
\usepackage{amsmath}
\usepackage{amssymb}
\usepackage{bm} % for \bm{}, better bold math than \boldsymbol

\begin{document}

\title{The Stellar Abundances and Galactic Evolution Survey (SAGES) V: The First Data Release of DDO51 Band}

\author[0000-0003-0278-7137]{Qiqian Zhang}
\affiliation{National Astronomical Observatories, Chinese Academy of Sciences, Beijing 100101,  People's Republic of China;  gzhao@nao.cas.cn; zfan@bao.ac.cn}
\affiliation{ School of Astronomy and Space Science, University of Chinese Academy of Sciences, Beijing 100049, People's Republic of China
}
\email{}

\author[0000-0002-6790-2397]{Zhou Fan}
\affiliation{National Astronomical Observatories, Chinese Academy of Sciences, Beijing 100101,  People's Republic of China;  gzhao@nao.cas.cn; zfan@bao.ac.cn}
\affiliation{ School of Astronomy and Space Science, University of Chinese Academy of Sciences, Beijing 100049, People's Republic of China
}
\email{zfan@nao.cas.cn}

\author[0000-0002-8980-945X]{Gang Zhao}

\affiliation{National Astronomical Observatories, Chinese Academy of Sciences, Beijing 100101,  People's Republic of China;  gzhao@nao.cas.cn; zfan@bao.ac.cn}
\affiliation{ School of Astronomy and Space Science, University of Chinese Academy of Sciences, Beijing 100049, People's Republic of China
}
\email{gzhao@nao.cas.cn}

\author[0000-0001-8424-1079]{Kai Xiao}
\affiliation{ School of Astronomy and Space Science, University of Chinese Academy of Sciences, Beijing 100049, People's Republic of China
}
\email{}

\author[0000-0002-9702-4441]{Wei Wang}
\affiliation{National Astronomical Observatories, Chinese Academy of Sciences, Beijing 100101,  People's Republic of China;  gzhao@nao.cas.cn; zfan@bao.ac.cn}
\email{}

\author[0009-0007-5610-6495]{Hongrui Gu}
\affiliation{National Astronomical Observatories, Chinese Academy of Sciences, Beijing 100101,  People's Republic of China;  gzhao@nao.cas.cn; zfan@bao.ac.cn}
\affiliation{ School of Astronomy and Space Science, University of Chinese Academy of Sciences, Beijing 100049, People's Republic of China
}
\email{}

\author[0000-0001-6637-6973]{Jie Zheng}
\affiliation{National Astronomical Observatories, Chinese Academy of Sciences, Beijing 100101,  People's Republic of China;  gzhao@nao.cas.cn; zfan@bao.ac.cn}
\email{}

\author[0000-0003-2868-8276]{Jingkun Zhao}
\affiliation{National Astronomical Observatories, Chinese Academy of Sciences, Beijing 100101,  People's Republic of China;  gzhao@nao.cas.cn; zfan@bao.ac.cn}
\email{}

\author[0009-0000-4835-7525]{Chun Li}
\affiliation{National Astronomical Observatories, Chinese Academy of Sciences, Beijing 100101,  People's Republic of China;  gzhao@nao.cas.cn; zfan@bao.ac.cn}
\email{}

\author[0000-0002-8442-901X]{Yuqin Chen}
\affiliation{National Astronomical Observatories, Chinese Academy of Sciences, Beijing 100101,  People's Republic of China;  gzhao@nao.cas.cn; zfan@bao.ac.cn}
\email{}

\author[0000-0003-2471-2363]{Haibo Yuan}
\affiliation{Department of Astronomy, Beijing Normal University, Beijing 100875, People's Republic of China}
\email{}

\author[0000-0002-0389-9264]{Haining Li}
\affiliation{National Astronomical Observatories, Chinese Academy of Sciences, Beijing 100101,  People's Republic of China;  gzhao@nao.cas.cn; zfan@bao.ac.cn}
\email{}

\author[0000-0003-0173-6397]{Kefeng Tan}
\affiliation{National Astronomical Observatories, Chinese Academy of Sciences, Beijing 100101,  People's Republic of China;  gzhao@nao.cas.cn; zfan@bao.ac.cn}
\email{}

\author[0000-0001-7255-5003]{Yihan Song}
\affiliation{National Astronomical Observatories, Chinese Academy of Sciences, Beijing 100101,  People's Republic of China;  gzhao@nao.cas.cn; zfan@bao.ac.cn}
\email{}

\author[0000-0001-7865-2648]{Ali Luo}
\affiliation{National Astronomical Observatories, Chinese Academy of Sciences, Beijing 100101,  People's Republic of China;  gzhao@nao.cas.cn; zfan@bao.ac.cn}
\email{}

\author[0000-0002-2241-7945]{Nan Song}
\affiliation{China Science and Technology Museum, Beijing 100101, People's Republic of China}
\email{}

\author[0009-0008-3430-1027]{Yujuan Liu}
\affiliation{National Astronomical Observatories, Chinese Academy of Sciences, Beijing 100101,  People's Republic of China;  gzhao@nao.cas.cn; zfan@bao.ac.cn}
\email{}

\author[0000-0002-8337-4117]{Yaqian Wu}
\affiliation{National Astronomical Observatories, Chinese Academy of Sciences, Beijing 100101,  People's Republic of China;  gzhao@nao.cas.cn; zfan@bao.ac.cn}
\email{}

\author[]{Ali Esamdin}
\affiliation{Xinjiang Astronomical Observatory, Urumqi 830011, People's Republic of China}
\email{}
\author[]{Hubiao Niu}
\affiliation{Xinjiang Astronomical Observatory, Urumqi 830011, People's Republic of China}
\email{}
\author[]{Jinzhong Liu}
\affiliation{Xinjiang Astronomical Observatory, Urumqi 830011, People's Republic of China}
\email{}
\author[]{Guojie Feng}
\affiliation{Xinjiang Astronomical Observatory, Urumqi 830011, People's Republic of China}
\email{}
\author[]{Yu Zhang}
\affiliation{Xinjiang Astronomical Observatory, Urumqi 830011, People's Republic of China}
\email{}

\begin{abstract}

We present the first public data release of DDO51 band from the Stellar Abundances and Galactic Evolution Survey (SAGES), based on Nanshan One-meter Wide-field Telescope (NOWT)  observations obtained between 2023 September and 2024 January. This release initiates the DDO51-band component of the survey, covering $\sim$ 2,500 deg$^2$ of the northern sky and including more than 10 million sources. The DDO51 filter is centered near the \ion{Mg}{1}~$b$ triplet  and the adjacent MgH feature, offering sensitivity to stellar surface gravity. The data reduction pipeline incorporates an improved astrometric solution anchored to Gaia DR3 and a photometric calibration strategy tied to synthetic photometry from Gaia XP spectra. These procedures yield a point-source depth of $\sim$18.9 mag at S/N$\sim$10 and an internal photometric precision $\approx$6-7 mmag at the bright end. A preliminary color--color analysis using Gaia broadband photometry confirms the expected sensitivity of the DDO51 band to stellar surface 
gravity, demonstrating a clear photometric separation between dwarf and 
giant sequences for late-type stars.  This dataset, when combined with existing SAGES photometry in other bands, provides a crucial tool for disentangling the substructures of the Milky Way. All data products from this release upon publication will be available.

\end{abstract}

% \keywords{\uat{Galaxies}{573} --- \uat{Cosmology}{343} --- \uat{High Energy astrophysics}{739} --- \uat{Interstellar medium}{847} --- \uat{Stellar astronomy}{1583} --- \uat{Solar physics}{1476}}

\section{introduction} 
\label{sec:intro}

The Gaia mission has transformed Galactic archeology by providing precise astrometry for over one billion stars \citep{prusti2016gaia, perryman2025space}.
However, to fully reconstruct the Milky Way's assembly history, this kinematic information must be complemented by precise stellar atmospheric parameters, including effective temperature ($T_{\rm eff}$), metallicity ([Fe/H]), and surface gravity ($\log g$). Such parameters are not yet available at comparable for the full Gaia sample. While spectroscopic surveys can provide the most accurate parameters, they are resource-intensive and generally limited to brighter targets.
Multiband photometry offers an efficient alternative, as color information across multiple passbands enables classifying stellar types and deriving fundamental atmospheric parameters \citep{bessell2005standard, casagrande2014synthetic}. 
Large-scale surveys such as the Sloan Digital Sky Survey (SDSS; \citealt{york2000sloan}) and Pan-STARRS \citep{kaiser2002pan,chambers2016pan} have revolutionized Galactic astronomy \citep{ivezic2008milky, ivezic2012galactic}.  Estimation of stellar atmospheric parameters from broadband photometry suffers from $T_{\rm eff}$–[Fe/H]–$\log g$ degeneracies, as different combinations of these parameters can produce similar broadband colors. in particular, distinguishing valuable halo giants from the overwhelming foreground of disk dwarfs remains challenging, as $\log g$ is only weakly constrained without dedicated gravity-sensitive diagnostics.

To break degeneracies and to obtain accurate stellar atmospheric parameters for large samples, the Stellar Abundances and Galactic Evolution Survey (SAGES; \citealt{wang2013stromgren,zheng2018sage,Fan2018sage,zheng2019sage}) was initiated.
SAGES is a multiband, deep photometric survey designed to cover $\gtrsim$ 12{,}000~deg$^{2}$ of the northern sky. It employs a unique eight-filter system tailored to target key diagnostic features in stellar spectra, consisting of $u_{\rm SC}$, $v_{\rm SAGES}$, DDO51, $g$, $r$, $i$, H$\alpha_{\rm n}$ and H$\alpha_{\rm w}$ filters. A detailed description of the filters is provided in Section~\ref{sec:Survey_Overview}.
Within this system, the intermediate-band DDO51 filter, centered on the gravity-sensitive Mg features, provides the primary photometric diagnostic for the surface gravity of G/K-type stars. 

The sensitivity of the Mg~I~b and MgH features to stellar surface gravity has been recognized for nearly a century \citep{ohman1936red, thackeray1939intensity}. The DDO51 filter was specifically designed to exploit these features, proving highly effective in separating late-type dwarfs and giants \citep{clark1979photoelectric}. Historically, \citet{geisler1984luminosity, geisler1990washington} demonstrated that pairing DDO51 with the Washington $M$ band creates a potent luminosity discriminator. This $(M - {\rm DDO51})$ color index subsequently became a cornerstone of Galactic archaeology. It has been extensively utilized to identify giant members in Milky Way dwarf spheroidals \citep{majewski2000exploring}, explore the Magellanic Clouds \citep{majewski2008discovery,nidever2011discovery}, and map the halo of M31 \citep{gilbert2012global, tollerud2012splash}. The APOGEE survey uses DDO51 photometry for its spectroscopic target pre-selection \citep{zasowski2013target}, and has proven very effective at identifying giant stars\citep{beaton2021final}.

Some studies have explored replacing the Washington $M$ band with alternative broadband filters.\citep[e.g.,][]{2007PhDT.........7T,casey2018infrared}.
\citet{2007PhDT.........7T} 
showed that the $V$ and $I$ bands can provide comparable 
dwarf-giant discrimination for relatively metal-rich populations.
However, metal-poor dwarfs can mimic metal-rich giants because the Mg absorption weakens at low metallicities \citep{morrison2001mapping}. These considerations highlight the importance of combining DDO51 with additional metallicity- and temperature-sensitive bands to mitigate residual degeneracies.

Within the SAGES filter system, DDO51 data will be combined with multiple filters, enabling a multi-dimensional approach rather than a single color–color diagram. Synthetic photometry analyses based on the SAGES filter set \citep{zhang2025stellar} indicate that the inclusion of DDO51 can substantially reduce the Teff–logg degeneracy inherent to broadband-only configurations. While the present work focuses on the data release and calibration of the DDO51 band, these results illustrate the expected role of DDO51 as the primary gravity-sensitive component of the SAGES system.

The primary science drivers for the SAGES DDO51 
observations include: (1) photometric pre-selection of giant 
stars for spectroscopic follow-up surveys; (2) large-scale, homogeneous determination of stellar atmospheric parameters ($T_{\rm eff}$, [Fe/H], $\log g$) across the northern sky; 
and (3) assisting with the photometric identification of metal-poor star candidates in combination with the $v_{\rm SAGES}$ or the other band.

The SAGES survey has been executed in phases using international facilities.
Observations for the $u_{\rm SC}$  and $v_{\rm SAGES}$ bands began in September 2015 using the 2.3 m Bok Telescope in the United States and continued until October 2019. Further progress was halted due to the global pandemic, leaving the observations for these two bands 87.8\% complete and reaching a completeness limit of $u_{\rm SC} \approx 20.4$ mag and $v_{\rm SAGES} \approx 20.3$ mag. These data formed the basis of the first SAGES data release (DR1; \citealt{fan2023stellar}), which was made public in June 2023.

To provide broadband optical coverage in SAGES regions not imaged by SDSS, complementary $g$, $r$, and $i$ observations were obtained between August 2016 and January 2018 using the Nanshan One-meter Wide-field Telescope (NOWT) in China.
These data cover $\sim 4{,}600~\mathrm{deg}^2$ and reach completeness limits of $g \approx 19.2$~mag, $r \approx 19.1$~mag, and $i \approx 18.2$~mag. 
They were presented in a supplementary release (DR1s; \citealt{li2024stellar}), which was made public in October 2024.

The facilities and schedules for the three remaining bands (DDO51, H$\alpha_{\rm n}$, and H$\alpha_{\rm w}$) have changed due to instrument availability and scheduling constraints. The DDO51 observations presented in this work began in September 2023 with the NOWT. Despite operational challenges affecting observing efficiency, including camera and shutter issues, substantial coverage has been achieved. Observations in the H$\alpha_{\rm w}$ band are currently planned with the Altay 1 m telescope.

The scientific utility of SAGES is already demonstrated by several recent studies. Notably, the unique photometric information from SAGES DR1 has been pivotal in large-scale stellar parameter estimation \citep[e.g.,][]{huang2023beyond,gu2025stellar}. The survey's sensitivity to metallicity has also been demonstrated by \citet{hong2024candidate}, who successfully used SAGES and SkyMapper data to identify metal-poor star candidates.

In this paper, we present the first release of observed DDO51-band photometry, covering $\sim 2{,}500$~deg$^{2}$. We describe the observations and data-reduction pipeline (Sections \ref{sec:observe}--\ref{sec:reduction}), the construction and content of the released catalog (Section \ref{sec:catalog}), the astrometric and photometric validation (Section \ref{sec:Validation}), and finally provide a science demonstration of dwarf–giant separation (Section \ref{sec:logg}). Section \ref{sec:access} describes how to access the SAGES DDO51 catalog. Section \ref{sec:summary} summarizes the main properties of DDO51 and outlines future work.

\section{Observations} 
\label{sec:observe}

\subsection{SAGES Survey Design and Overview}
\label{sec:Survey_Overview}
SAGES targets the northern sky with $ \delta  > -5^\circ$, explicitly excluding the crowded, high-extinction Galactic plane region with
$|b|< 10^\circ$. In addition, motivated by the site’s seasonal observing conditions, we restricted the survey to fields observable in autumn–winter and excluded fields with $180^\circ < $RA $<270^\circ$. The final footprint of the survey spans approximately $12{,}000~\mathrm{deg}^2$, shown as the gray background in Figure~\ref{fig:sky_coverage}.

\begin{figure}
    \centering
    \includegraphics[width=0.9\linewidth]{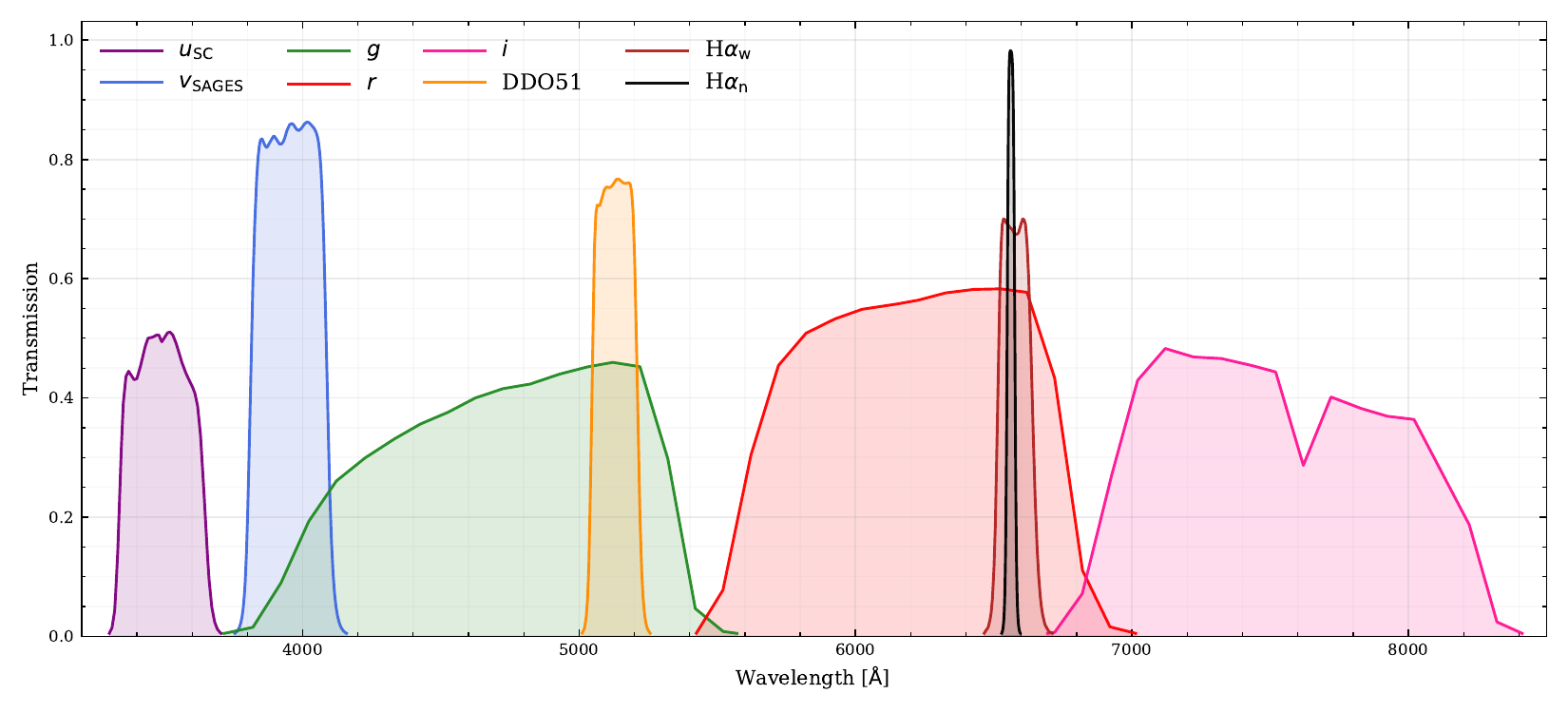}
    \caption{Total system throughput curves of the SAGES passbands, including the filter transmission and CCD quantum efficiency. The H$\alpha_n$ filters are shown using the design transmission profiles, 
    while the other filters are based on laboratory measurements.} 
    \label{fig:filter_plot}
\end{figure}

The survey employs eight filters designed to capture key stellar-atmosphere diagnostics: the medium-band Str\"{o}mgren--Crawford $u_{\rm SC}$ filter(Balmer jump; surface gravity for early-type stars); the SAGES-designed $v_{\rm SAGES}$ filter(\ion{Ca}{2} H\&K; metallicity sensitivity); The DDO51 band, at a center wavelength of 5130~\AA\ with a FWHM of $\sim 162$~\AA\, is covered on gravity sensitive Mg b features (dominated by \ion{Mg}{1} 5167/5173/5184 \AA) and the adjacent MgH band. This spectral feature exhibits pronounced pressure-broadened wings that are highly sensitive to the stellar surface gravity of GK-type stars (see Figure~1 of \citealp{zhang2025stellar}). SDSS-like broad $g$, $r$, and $i$ bands; and the narrow- and wide-band H$\alpha_{\rm n}$ and H$\alpha_{\rm w}$ filters, which probe H$\alpha$ line strength and, in combination with broadband colors, help constrain extinction and activity/emission-line objects. The central wavelengths and bandwidths, , and the current data release status are summarized in Table~\ref{tab:SAGES_filters}; the full passband curves are shown in  Figure \ref{fig:filter_plot}.

\begin{deluxetable}{lcccccccc}
\tablecaption{SAGES Filter Parameters\label{tab:SAGES_filters}}
\tablewidth{0pt}
\tablehead{
\colhead{\textbf{Filter}} & \colhead{$u_{\rm SC}$} & \colhead{$v_{\rm SAGES}$} & \colhead{$g$} & \colhead{$r$} & \colhead{$i$} & \colhead{H$\alpha_{\rm n}$} & \colhead{H$\alpha_{\rm w}$} & \colhead{DDO51}
}
\startdata
\textbf{Central Wavelength (\AA)} & 3425 & 3950 & 4686 & 6166 & 7480 & 6563 & 6563 & 5130 \\
\textbf{Bandwidth (\AA)}          & 314  & 290  & 1280 & 1150 & 1230 &  29  & 136  &  162 \\
\textbf{Observation Status}      & Done  & Done  & Done & Done & Done &  Planned  & Planned  &  In progress \\
\textbf{Release Status}  &DR1\tablenotemark{a}  & DR1\tablenotemark{a}  & DR1s\tablenotemark{b} & DR1s\tablenotemark{b} & DR1s\tablenotemark{b} &  --  & --  &  This work (partial) \\
\textbf{Sky coverage (deg$^2$)} & $\sim$9960 & $\sim$9960 & $\sim$4600 & $\sim$4600 &$\sim$4600 & -- & -- & $\sim$2500 \\
\enddata
\tablenotetext{a}{Released in SAGES Data Release 1  \citep{fan2023stellar}}
\tablenotetext{b}{Released in SAGES DR1 supplementary release \citep{li2024stellar}}

\end{deluxetable}

Balancing the constraints of the telescope aperture, exposure time limits, and system efficiencies, we set design target depths of S/N $\approx 100$ at $u_{\rm SC} \approx 17$~mag, $v_{\rm SAGES} \approx 16$~mag, and $\approx 15$~mag in the remaining bands ($g$, $r$, $i$, DDO51, H$\alpha_{\rm n}$, and H$\alpha_{\rm w}$).  
These depth goals ensure that SAGES can obtain high-quality photometry for large samples of stars across the main-sequence and giant branches, providing robust constraints on $T_{\rm eff}$, [Fe/H], and $\log g$ over the full survey footprint.

\subsection{Facility}
All DDO51 observations presented in this release were acquired using the NOWT located at Nanshan Station $(87^\circ 10^{\prime}\,\mathrm{E},\; 43^\circ 28^{\prime}\,\mathrm{N},\; 2080~\mathrm{m})$\citep{bai2020wide,shan2021photometry,li2024stellar}. Typical seeing at the site is better than ~$2.2\arcsec$.
The NOWT is a prime-focus system ($f/2.2$) equipped with a field derotator to maintain alignment with celestial coordinates during exposures. The system features a $1000$ mm effective aperture parabolic primary mirror with an effective focal length of $\sim 2{,}160$~ mm. 

The imaging camera is a liquid-nitrogen-cooled, blue-sensitive E2V CCD with a resolution of $4096\times 4136$ pixels. The $12\,\mu\mathrm{m}$ pixel size yields a plate scale of $1\farcs146~\mathrm{pixel}^{-1}$, resulting in an effective field of view (FoV) of $1.3^{\circ}\times 1.3^{\circ}$. The detector utilizes a four-amplifier readout system with a 16-bit A/D converter. Each amplifier includes a 32-pixel overscan strip used to track the bias level. The quantum efficiency at the DDO51 central wavelength ($\sim 5130$~\AA) is approximately 80\%. For the DDO51 survey component, the exposure time was fixed at 180~s to achieve the target depth.

\subsection{The DDO51 Band Dataset}
\label{sec:ddo51_overview}
\begin{figure}
    \centering
    \includegraphics[width=0.7\linewidth]{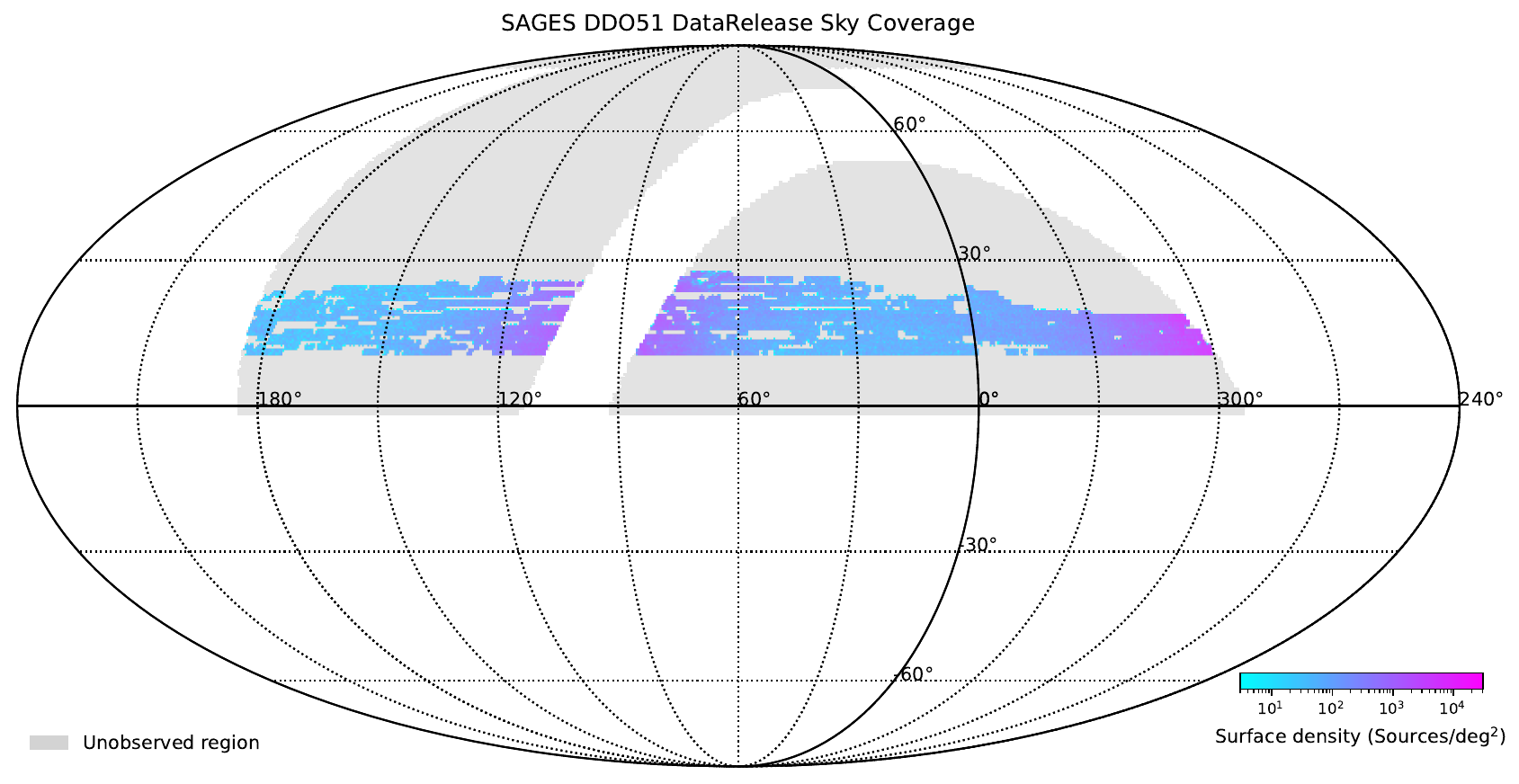}
    \caption{Sky coverage of the SAGES DDO51 data released in this work. The gray region shows the full SAGES survey footprint, while the colored pixels mark the area included in this release, with color indicating the surface density of detected sources (sources per deg$^{-2}$).}
    \label{fig:sky_coverage}
\end{figure}

To minimize sky background contamination from moonlight, observations were restricted to a window of $\pm 5$ days around the New Moon, with a further requirement that the target's angular separation from the Moon exceed $50^\circ$ and the separation from the Sun exceed $60^\circ$. Accounting for CCD readout overheads and telescope slewing time, the operational cycle for each exposure was set to 240~s. Furthermore, to minimize airmass during the observing window, the survey strategy prioritized higher altitudes, commencing at $\mathrm{Dec} \ge +10^\circ$ and scanning progressively northward.

This data release comprises DDO51 observations obtained under these conditions between September 2023 and January 2024. A total of 41 nights were allocated for these observations, with actual data acquisition occurring on 29 nights due to weather constraints and instrument maintenance. The raw dataset consists of 2,896 images, which were reduced to 2,242 high-quality science images after quality control filtering (see Section \ref{sec:reduction}). The resulting sky coverage is shown in Figure~\ref{fig:sky_coverage}, spanning approximately 2,500 deg$^2$. The color map indicates the surface density of detected sources, highlighting the survey's coverage density.

\subsection{Observing Strategy}
\begin{figure}
    \centering
    \includegraphics[width=0.5\linewidth]{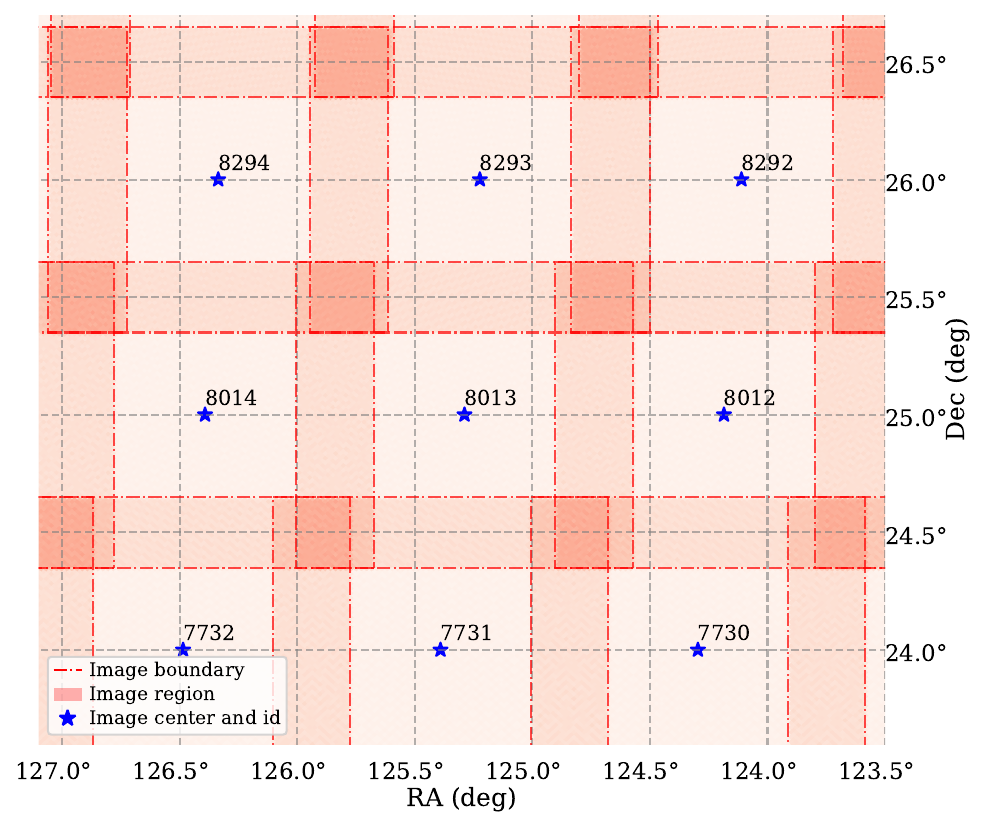}
    \caption{Example of the SAGES pointing pattern in equatorial coordinates. Red dashed rectangles outline the individual image footprints, with shaded regions indicating the effective image area and their mutual overlaps, while blue stars mark the image centers labeled by field ID. }
    \label{fig:tiling}
\end{figure}

The survey footprint is partitioned into declination-fixed strips. Adjacent field centers are separated by $1^\circ$, providing deliberate overlaps that (i) secure a dense network of cross-matched sources for astrometric and photometric cross-calibration, (ii) enable stacking for improved signal-to-noise, and (iii) accommodate modest telescope pointing errors while preserving high completeness across the survey footprint. Figure~\ref{fig:tiling} illustrates the tiling pattern in a representative region of the sky, where the mutual overlaps between adjacent pointings are clearly visible.

We developed an observation scheduler that automatically generates nightly observing scripts under realistic operational constraints \citep{zheng2024strategies}. Given a site/telescope configuration and a date, the scheduler will evaluate each to-be-observed region’s visibility, altitude/airmass, and angular separations from the Moon and Sun, and select the best fields of each time window. Subject to predefined constraints—including global declination limits, hour-angle limits, an airmass ceiling, a maximum-altitude cap (to avoid near-zenith operations), and solar/lunar avoidance radii—the algorithm ranks candidate blocks by lower airmass, smaller absolute hour angle (meridian proximity), and lower declination, while penalizing slews relative to the previous pointing with an asymmetric cost that suppresses large westward moves, thereby minimizing large telescope repointings between regions.

\section{Data Reduction} 
\label{sec:reduction}

\subsection{Real-time Quality Control}
Real-time quality control (QC) ensures that only images meeting the design image-quality and depth requirements enter the data-reduction pipeline. We have established a real-time QC system that integrates automated filtering with human supervision, allowing both rapid response and contextual judgment. The observer monitors the real-time quality reports generated by the automated system. The automated system performs immediate quantification of each image's quality and the automatic flagging of substandard data. 

Images are rejected if affected by (1) adverse atmospheric conditions (e.g., high humidity, cirrus, turbulence), (2) telescope tracking errors, (3) electronic noise issues, and (4) derotator malfunctions. Such contaminated frames would fail to achieve the designated limiting magnitude, thereby degrading the photometric completeness and overall quality of the final catalog. Consequently, non-compliant images are excluded, and their corresponding sky regions are scheduled for future re-observation.

Our automated filtering system runs as a background daemon on the data reduction server. Following the completion of each exposure, the daemon executes a rapid processing routine, typically completed within several tens of seconds per image. This process first invokes Source Extractor (\citealt{bertin1996sextractor}) on each raw quick-look image to extract sources. From these, preliminary quality metrics are obtained.

The system quantifies two key parameters from  Source Extractor, FWHM\_IMAGE and ELONGATION. {FWHM\_IMAGE represents the Full-Width at Half-Maximum of the object's profile, which Source Extractor estimates by assuming a Gaussian core at both global and local levels. Only unflagged (Source Extractor's \texttt{FLAGS}=0) sources are used in the computation of quality metrics. Globally, median values across the entire field are computed, while locally, the image is divided into a $10\times10$ grid to assess spatial variations. This local approach is particularly effective for identifying issues like tracking jumps or derotator malfunctions, which often impact only localized regions.

We define the acceptance criteria for an image as follows. First, the global median FWHM must be less than 5.0 pixels, and the global median ELONGATION must be less than 1.4. Second, at least 95\% of the $10\times10$ local grid cells must also satisfy a median FWHM not exceeding 7.0 pixels and a median ELONGATION not exceeding 1.7. These criteria were empirically determined from calibration observations to balance sensitivity and robustness.

All derived quality metrics are visualized in real-time on the observation monitor. Images that fail any of the defined thresholds are immediately flagged as non-compliant and excluded from downstream processing. Simultaneously, a warning is displayed to alert the observer, prompting them to investigate potential issues with the telescope or weather and take corrective actions. Furthermore, the observer manually inspects each image when feasible, examining all image regions for anomalies that the automated Source Extractor-based detection might miss, thereby providing an additional layer of quality assurance. This combination of automated detection and manual inspection ensures comprehensive real-time quality assurance across all observing conditions.
\begin{figure*}
\gridline{
  \fig{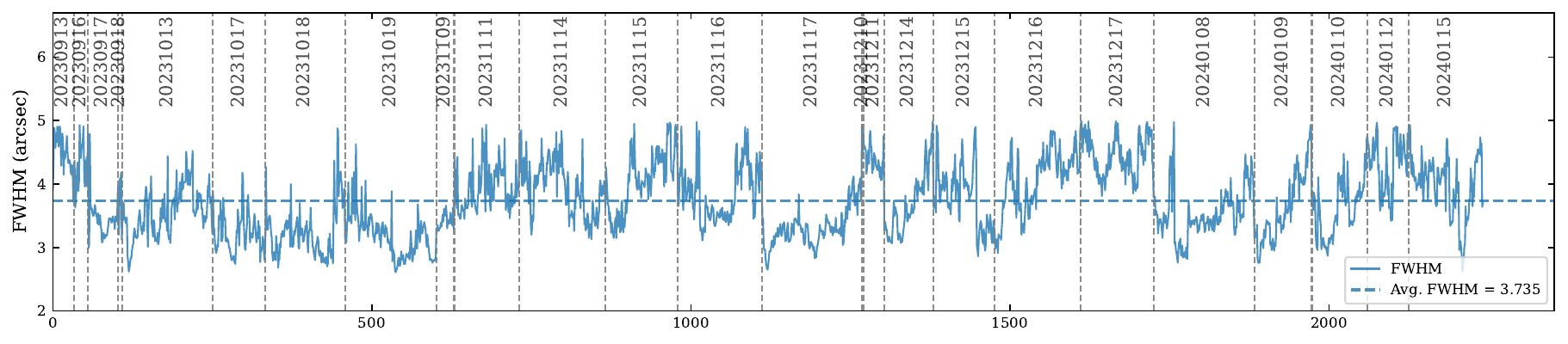}{0.9\textwidth}{(a) Median FWHM versus exposure index}
  }
  \gridline{
  \fig{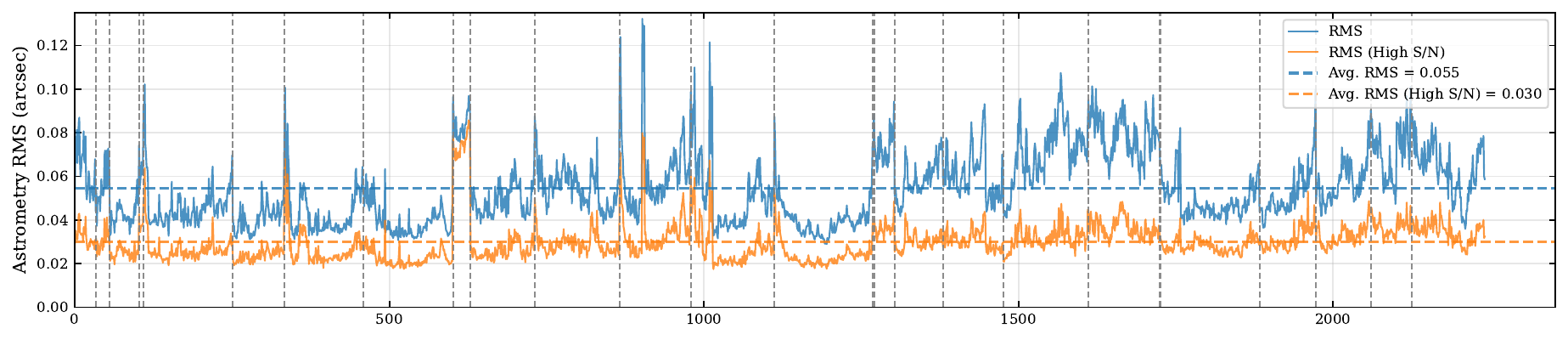}{0.9\textwidth}{(b) Astrometric RMS output by SCAMP versus exposure index.}
  }
  \gridline{
  \fig{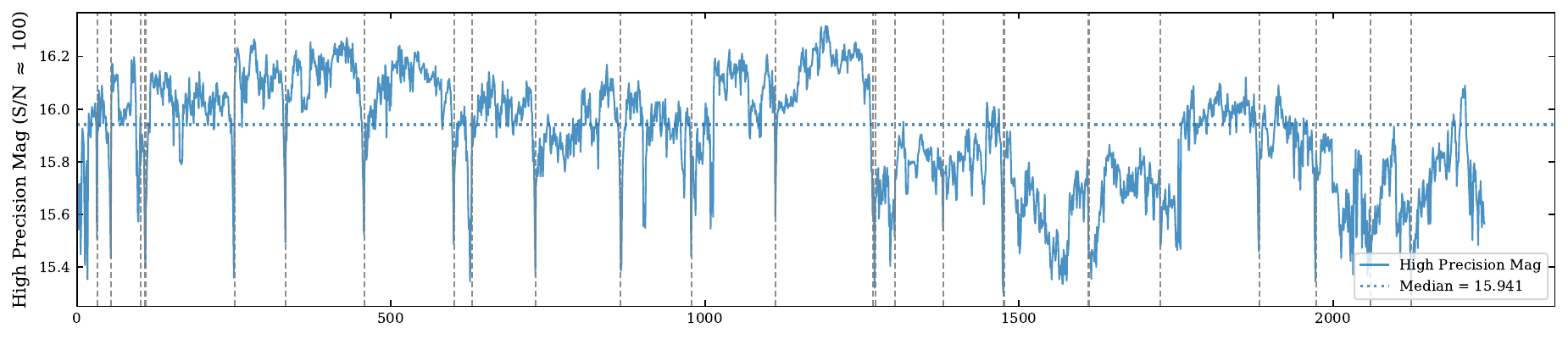}{0.9\textwidth}{(c) Magnitude at $\sigma\approx0.01$ mag (proxy for S/N$\approx$100) versus exposure index.}}
  \gridline{
  \fig{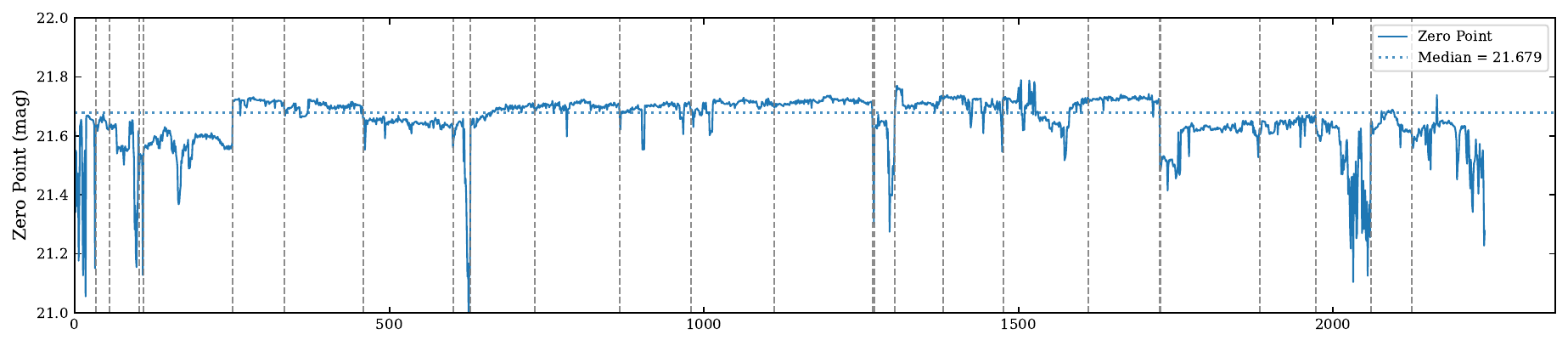}{0.9\textwidth}{(d) Photometric zero point versus exposure index.}}

\caption{
Long-term stability of image quality, astrometry, and photometry. 
All panels show image-level quantities as a function of exposure index, which follows the chronological observing sequence.
\textbf{(a)} Median FWHM for each exposure, tracing the temporal behavior of the seeing; vertical dashed lines mark the boundaries between observing nights.
\textbf{(b)} External astrometric RMS from SCAMP, both for all matched sources and for a high-S/N subsample, together with their median values, quantifying astrometric solution stability.
\textbf{(c)} The magnitude where the median formal uncertainty equals 0.01 mag ($\approx1$\% precision) and its evolution with time.
\textbf{(d)} Photometric zero point per exposure, with its median value overplotted, illustrating temporal stability and slow drifts in photometric calibration.
In all panels, the horizontal dashed line denotes the median value over the full sample of exposures.
}
\label{fig:stability}
\end{figure*}

\subsection{Image Processing and Calibration}
For each observing night, in addition to science exposures, of bias frames and twilight (evening or morning) sky flats are obtained. Prior to any subsequent calibration, all images undergo an overscan correction. For each amplifier, the median level of the overscan region is computed row by row to form a one-dimensional bias vector, which is then smoothed and subtracted from the corresponding rows of the science image before trimming the overscan region. All bias frames are then combined on a pixel-by-pixel basis using a median to produce a master bias frame, from which the camera’s readout noise is estimated and recorded.

Flat fielding requires particular attention due to the four amplifier readout architecture, each with its own analog chain and A/D converter. The conversion gain ($\mathrm{e}^-\,\mathrm{ADU}^{-1}$) and photometric zero-points of these channels differ slightly and may drift gradually or discontinuously over time. To prevent such variations from introducing normalization errors or photometric zero-point biases, each flat image—after overscan and bias subtraction—is first normalized per amplifier so that the mean level of each readout region equals unity. All normalized flats are then median-combined pixel by pixel to construct the master flat, which is stored as a multi-extension FITS (MEF) file, with one image extension per amplifier. This procedure mitigates the impact of gain drifts and ensures balanced weighting among individual frames.

The standard preprocessing sequence for science images consists of three steps: overscan correction, subtraction of the master bias frame, and division by the normalized master flat. Finally, fluxes are normalized to a $1~\mathrm{s}$ exposure time.

Source extraction is performed using the widely used astronomical software Source Extractor. It identifies and measures sources in the processed images, providing instrumental magnitudes and pixel coordinates. Consistent with DR1/DR1s, we adopt \texttt{MAG\_AUTO} as the instrumental magnitude, which implements a flexible, Kron-like elliptical aperture, that dynamically scales and rotates to match the local seeing conditions, optical distortions, and the source's intrinsic profile across the wide field of view.

We opted for this adaptive aperture photometry over Point-Spread-Function (PSF) fitting for three primary reasons: (1) it maintains consistency with prior SAGES releases (DR1 and DR1s); (2) our footprint strictly avoids the crowded Galactic plane ($|b| > 10^\circ$), making aperture photometry highly reliable for our predominantly isolated sources; and (3) \texttt{MAG\_AUTO} is inherently robust against spatially varying PSFs, which is highly dependent on accurate PSF modeling.

The CCD array is exceptionally clean, with zero dead pixels (response $< 0.1$) and only 83 pixels out of 16 million exhibiting a response below 0.7. To rigorously handle these specific defects, a static bad pixel mask ($< 0.7$ response) is supplied to Source Extractor. Any source whose aperture overlaps a masked pixel triggers an internal warning, which is propagated into our catalog as \texttt{FLAG\_SE\_FLAGS} (Section~\ref{sec:catalog}) and rejected during the catalog merging process.

\subsection{Astrometric Calibration}
The astrometric calibration for this data release significantly improved upon the procedures used for DR1/DR1s. A key improvement is the transition from the PPMX
catalog \citep{roser2008ppm} to the Gaia DR3 catalog \citep{2023A&A...674A...1G} as the fundamental astrometric reference frame. With sub-milliarcsecond accuracy for bright stars and precise proper motions, Gaia DR3 allows for highly accurate coordinate propagation to the common epoch of J2023.8, enabling a robust and precise World Coordinate System (WCS) solution.

Our astrometric calibration pipeline consists of two main stages: a coarse solution and a subsequent refined fitting procedure.

\subsubsection{Initial Field Identification}
For each science image, we first perform a preliminary source extraction. The top 1,000 brightest, unsaturated sources are then processed with Astrometry.net\footnote{\url{https://nova.astrometry.net/}} \citep{lang2010astrometry} to obtain a coarse WCS solution. The primary purpose of this step is to robustly determine the pointing center and rotation of the image and to provide a first-order mapping between pixel and celestial coordinates, which is then used in the subsequent stage.

\subsubsection{Refined WCS fitting with SCAMP}
The refined astrometric calibration is derived using SCAMP \citep{bertin2006automatic} in a two-round iterative scheme to improve robustness and accuracy. First, using the pointing information from Astrometry.net, we query a circular region in the Gaia DR3 catalog centered on the image, and apply a magnitude cut of 
$12 < G < 18$ mag. This range effectively removes saturated bright stars and low–S/N faint stars, ensuring that only high-quality sources are used for matching.

A crucial pre-processing step is the propagation of all Gaia reference positions from their native epoch (J2016.0) to the common epoch of observation (J2023.8) using their full 5-parameter propagation including parallax; perspective terms are included when RV is available. This correction is performed using \texttt{PyGaia} ensure strict epoch consistency.

Following the same methodology as in DR1/DR1s, SCAMP is run twice. In the first round, a larger matching radius is used to robustly cross-match the instrumental source catalog against the propagated Gaia reference, yielding a corrected, intermediate WCS solution. In the second round, a much tighter matching radius is applied, using the intermediate solution as input. This second iteration refines the fit and solves for a fifth-order polynomial distortion model using the SIP (Simple Imaging Polynomial) convention. The final high-precision WCS is then written into the FITS header of each image, ensuring a consistent and accurate astrometric solution across the entire survey.

\begin{figure}
    \centering
    \includegraphics[width=0.5\linewidth]{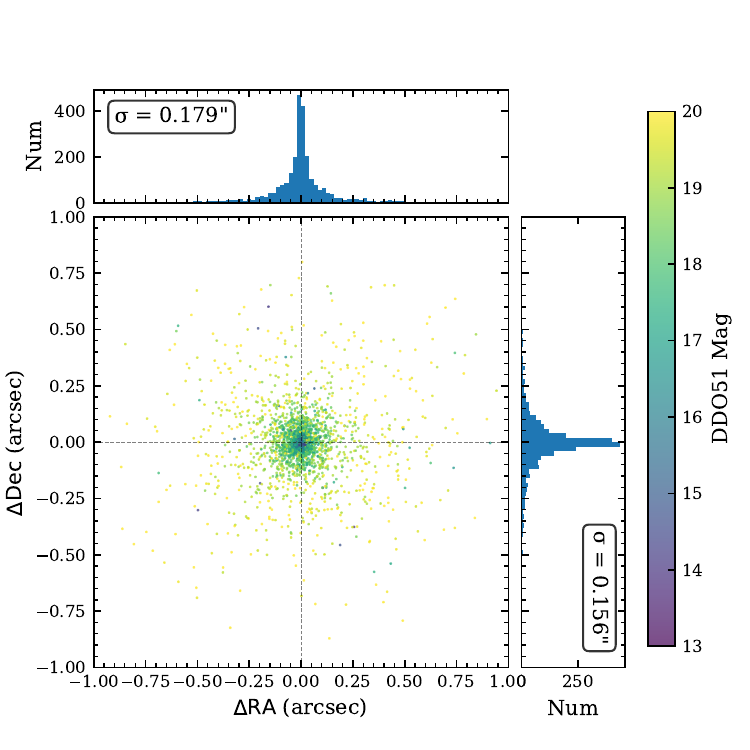}
    \caption{
        Two-dimensional distribution of astrometric residuals ($\Delta$RA vs.\ $\Delta$Dec)
        for one typical image, with points color-coded by DDO51 magnitude.
        The top and right panels show the one-dimensional histograms of $\Delta$RA and $\Delta$Dec,
        with the quoted $\sigma$ values marking the dispersion of each component.
    }
    \label{fig:radec_rms}
\end{figure}

\subsection{Cross-matching with Gaia DR3}
To achieve high-precision association between the observed sources and the Gaia DR3 reference catalog, we implemented a probabilistic cross-matching pipeline based on positional error ellipses of both observed sources and Gaia candidates. We identified the nearest Gaia source within a $5\sigma$ threshold and angular separation $<1\arcsec$ as the valid match. This approach accounts for geometric distortions, centroiding uncertainties, systematic astrometric residuals, and the epoch propagation errors of reference stars.

\subsection{Photometric Calibration}

The SAGES DR1 and DR1s data sets adopted the spectroscopy based stellar color regression method (SCR method) \citep{yuan2015stellar,huang2022photometric} and the photometric-based SCR method (SCR' method)\citep{xiao2023} for photometric calibration. This approach utilizes spectroscopic data from LAMOST and photometry from Gaia to predict the intrinsic colors of stars, thereby generating a set of color standards for calibration. However, the effectiveness of this method is constrained by the sky coverage of spectroscopic data and photometric zero-point transfer between different images, leading to a degradation in calibration precision.

The advent of Gaia DR3 has provided a powerful resource: 220 million absolutely calibrated, low-resolution XP spectra covering the entire sky \citep{2023A&A...674A..33G}, and comprehensive correction by \cite{huang2024comprehensive}. We reconstruct the Gaia XP spectra into flux as a function of wavelength using the published coefficients, obtaining a calibrated $f_{\lambda}(\lambda)$ spectrum for each source. The reconstructed flux is then convolved with the DDO51 corresponding filter passband response to compute the band-integrated flux, from which the synthetic magnitude is derived. We can synthesize highly accurate DDO51-band magnitudes, which serve as a dense and precise grid of standard stars.

The photometric calibration of the SAGES DDO51 survey data aims to achieve millimagnitude-level precision. A comprehensive description of the methodology, validation, and error analysis will be presented in a dedicated paper (Xiao et al. 2026, in preparation). 
% Our calibration pipeline is anchored entirely to the BEst STar (BEST) database, which is constructed from Gaia XP spectra. The BEST database provides an all-sky, high-precision catalog of standard stars in the SAGES DDO51 band, thus firmly tying our photometry to the Gaia photometric system.

To accurately correct for complex systematic effects arising from the four-amplifier CCD, the optical system, and the atmosphere, we have developed a two-step calibration strategy.

First, we construct a physical calibration model that explicitly accounts for inter-amplifier gain variations, which employs a two-dimensional polynomial to describe the zero-point offset as a function of detector pixel coordinates. For a given star in a single exposure, we model the magnitude difference, $\Delta m \equiv m_{\rm XPSP}-m_{\rm inst}$, where $m_{\rm XPSP}$ is the reference magnitude synthesized from Gaia XP spectra and $m_{\rm inst}$ is the instrumental magnitude. The spatially coherent, field-wide component of $\Delta m$ is described by a two-dimensional polynomial in the detector pixel coordinates $(x,y)$, capturing smooth sensitivity variations across the focal plane. This will also correct the center-to-edge filter bandpass shift caused by the varying angle of incidence on the interference filter. In addition, we include an independent, amplifier-specific zero-point term for each readout channel ($i=0,1,2,3$) to account for channel-to-channel gain differences. This per-amplifier constant directly quantifies and corrects the inter-amplifier gain variations, which are typically at the few-percent level, with occasional larger night-to-night jumps reported in the data. Overall, this calibration step simultaneously corrects inter-amplifier gain offsets and large-scale flat-field structures (e.g., vignetting).In addition, after correcting the position-dependent terms, we apply a per-exposure linear magnitude-dependent adjustment to remove weak residual trends with source brightness, and iterate this adjustment together with the spatial model until convergence.

Second, after applying the analytical model, we identified residual, medium-scale flat-field structures that cannot be described by a low-order polynomial. Through analysis of images of the same field taken at different de-rotation angles, we have confirmed that these medium-scale structures originate primarily from the telescope's optical system rather than the detector itself. We therefore apply a data-driven stellar flat-field correction. For each target source, this method utilizes the residuals ($m_{\text{XPSP}} - m_{\text{cal\_step1}}$) of its 80 nearest standard stars on the detector plane to fit a local linear correction (a sliding linear kernel) as a function of detector position, enabling the precise removal of these medium-scale patterns. 
\begin{figure}
\gridline{
  \fig{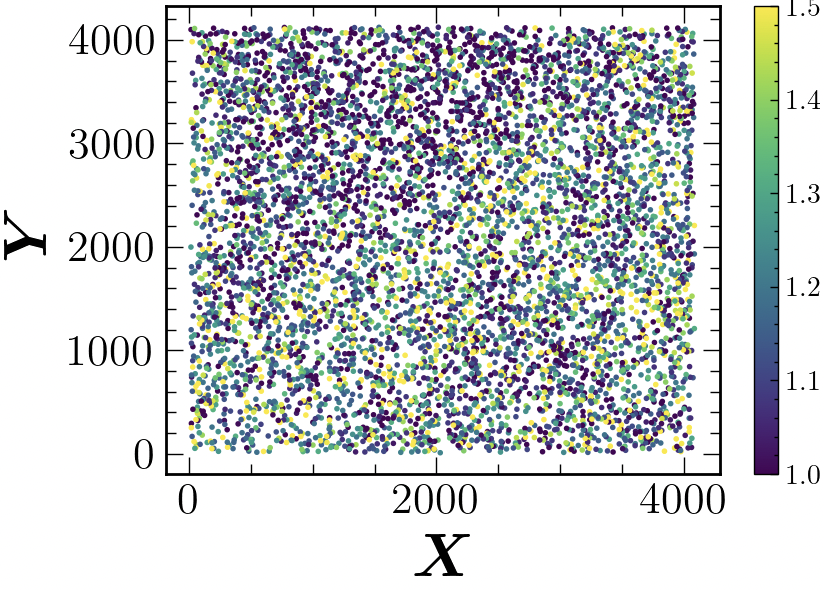}{0.41\textwidth}{(a)}
  \fig{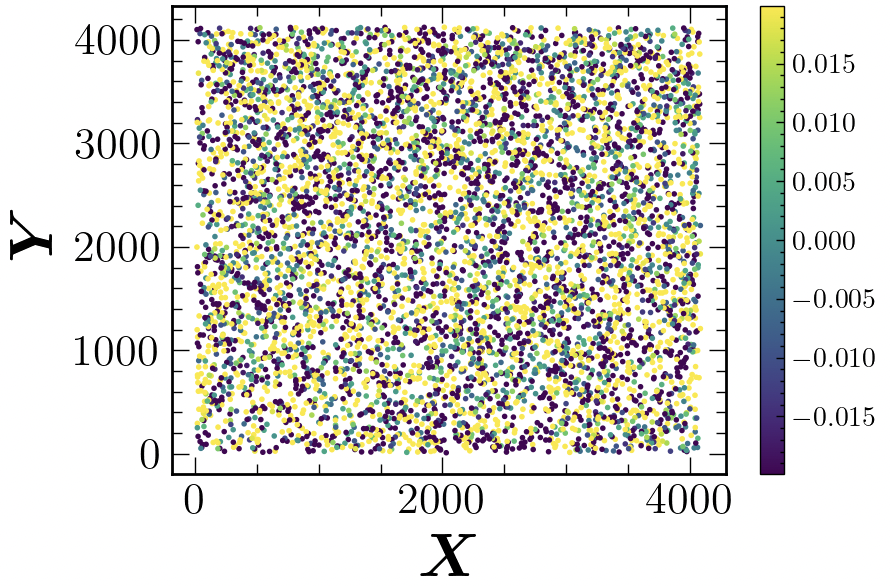}{0.45\textwidth}{(b)
}
}
\caption{Visualization of the 2D spatial photometric calibration for a representative single exposure. 
(a) The distribution of the reference standard sources (synthesized from Gaia DR3 XP spectra) across the detector pixel coordinates (X, Y), color-coded by their intrinsic Gaia $BP-RP$ colors.  
(b) The spatial distribution of the photometric residuals ($\Delta m = m_{\rm calibrated} - m_{\rm XP}$) after applying the full 2D spatial calibration and amplifier-specific zero-point corrections.}  
\label{fig:res}
\end{figure}

To visualize the effectiveness of this spatially resolved photometric calibration, Figure~\ref{fig:res} presents the spatial distribution of the standard stars and the corresponding calibration residuals for a representative single exposure. As shown in Figure~\ref{fig:res}(a), the dense and uniformly distributed grid of XP synthetic standard stars, covering a wide range of intrinsic colors. After applying full corrections, the resulting photometric residuals ($\Delta m = m_{\rm calibrated} - m_{\rm XP}$) exhibit a flat and near-zero distribution across all detector coordinates, as illustrated in Figure~\ref{fig:res}(b). This confirms that our calibration pipeline successfully characterizes and removes both detector-level non-uniformities and the intrinsic, spatially dependent optical signatures of the interference filter.
Figure~\ref{fig:mag_magerr} shows the relation between calibrated DDO51 magnitude and the photometric uncertainty reported by Source Extractor (\texttt{MAGERR\_AUTO}). 

Validation of the calibrated photometry, including internal precision from overlapping observations and external accuracy checks against independent references, is presented in Section~\ref{sec:Validation}. Additional methodological details and extended tests will be described in Xiao et al. (2026, in preparation).

\begin{figure}
    \centering
    \includegraphics[width=0.5\linewidth]{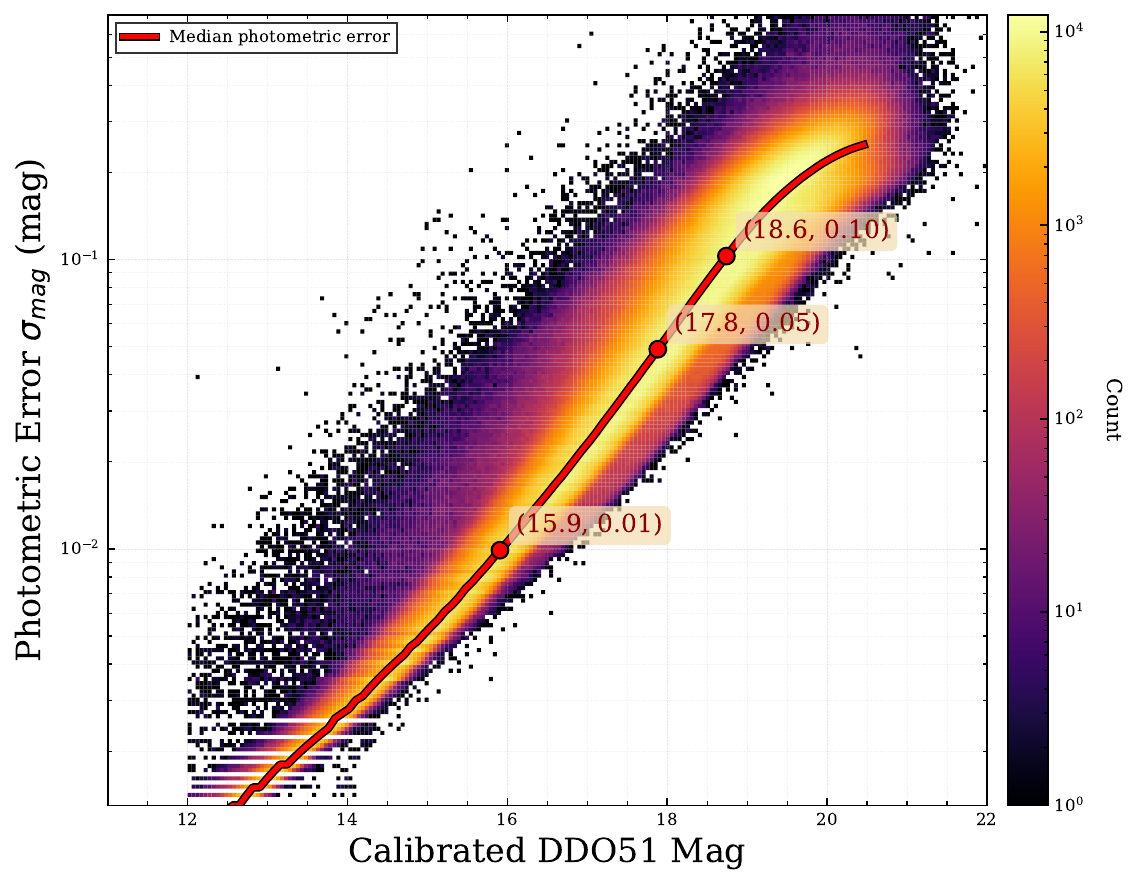}
    \caption{Calibrated DDO51 magnitude versus photometric uncertainty (\texttt{MAGERR\_AUTO} from Source Extractor) from the full single-epoch catalog (without catalog merging) after photometric calibration. }
    \label{fig:mag_magerr}
\end{figure}
\subsection{Identification and Removal of Linear Artifacts}

Ground-based optical surveys are increasingly affected by artificial satellites and other transient linear features, such as passing aircraft, meteor trails, which traverse the field of view during exposure. Charge blooming from saturated bright stars can also leave linear trails on the CCD frame. In addition, sporadic electronic issues—such as grounding failures that introduce inter-amplifier cross-talk—can cause significant noise spikes during the readout of specific rows, generating spurious detections. When processed by Source Extractor, all these linear traces on the image are fragmented into multiple detections, resulting in groups of false sources that appear as near-linear distributions in the extracted source catalogs.

To identify and eliminate such artifacts, we implemented an iterative RANSAC-based (Random Sample Consensus, \citealt{fischler1981random}) algorithm specifically designed to detect linear trajectories among sources that lack a reliable counterpart in the Gaia catalog (defined as having no match within a $10\arcsec$ radius). This pre-selection effectively removes genuine stars from the input sample and focuses the search on the population of unmatched detections, where spurious streak-like sources are most likely to occur. 

RANSAC provides a robust model-fitting framework that can recover the dominant linear structure in data heavily contaminated by outliers (in this case, true astrophysical sources). In each iteration, the algorithm randomly selects two points to define a candidate line and evaluates all remaining points by their orthogonal distances:
\begin{equation}
D = \left\| (p_i - p_0) - \left((p_i - p_0) \cdot d\right) d \right\| \leq \tau
\end{equation}
where $p_i$ is the point being tested, $p_0$ is a point on the line, $d$ is the unit direction vector, and $\tau$ is the residual (distance) threshold. Points satisfying this criterion are regarded as inliers. The process repeats for a large number of random trials, and the line model yielding the largest inlier set is retained as the best candidate.

To refine the preliminary RANSAC solution, we apply principal component analysis (PCA) to the identified inliers, which provides a total-least-squares estimate of the line parameters and ensures numerical stability even for nearly vertical trajectories. Once a valid track is identified, its inlier points are removed from the sample, and RANSAC is re-executed on the remaining data. This iterative procedure continues until no statistically significant linear structure remains.

To improve robustness, we apply several post-processing steps to the raw RANSAC solutions. 
First, to prevent separate tracks from being spuriously merged across gaps, we sort inlier points along the fitted principal direction and split a candidate trajectory into multiple segments whenever the separation between adjacent points exceeds a maximum allowed gap (300 pixels). 
Second, we reject overly sparse segments by requiring the typical inter-point spacing along the track to be smaller than a threshold (200 pixels); segments that fail this criterion are treated as false positives. 
For each accepted segment, we estimate its endpoints from the extrema of the projected coordinates, and then flag (\texttt{FLAG\_FAKE\_SOURCE=True}) as spurious all detections in the full source catalog that fall within a fixed-width corridor (10 pixels) around the fitted line.

This approach combines geometric interpretability, strong resistance to outliers, and is computationally efficient for large-scale surveys. It greatly reduces satellite and other linear contamination from wide-field survey data, producing a cleaner and more reliable source catalog for subsequent photometric calibration and astrophysical analysis.
\begin{figure}
\gridline{
  \fig{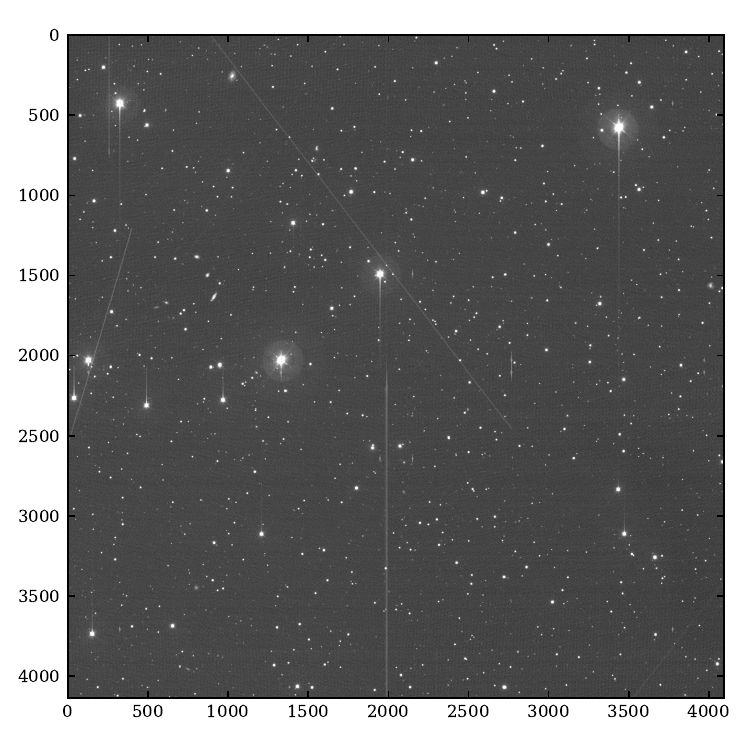}{0.45\textwidth}{(a) Original science image with multiple linear streaks}
  \fig{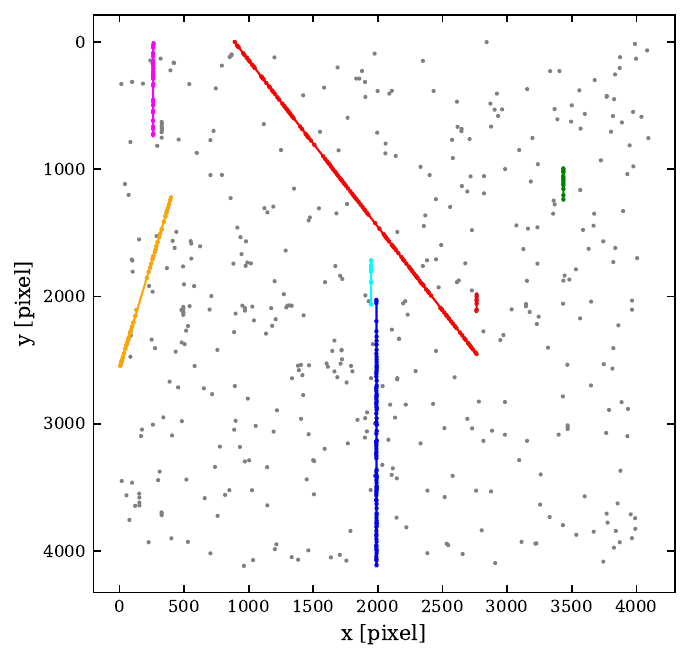}{0.45\textwidth}{(b) Unmatched detections and RANSAC-identified linear artifacts.}
}
\caption{
Left: example science exposure showing multiple linear streaks across the field. 
Right: sources that fail to match Gaia in the same image, plotted in pixel coordinates; 
colored segments mark spurious linear features identified as artifacts by the RANSAC-based detection algorithm.
}
\label{fig:ab}
\end{figure}

\section{Catalog Construction}
\label{sec:catalog}
\subsection{Source FLAGS}
To ensure the reliability of the final catalog and facilitate downstream scientific analyses, we established a comprehensive system of quality flags. This system is designed to identify and annotate sources that may be affected by instrumental artifacts, poor measurement quality, or environmental contamination.

Each source in the catalog carries a bitmask column named \texttt{FLAGS\_SAGES}. If a source meets any of the following criteria, the corresponding binary bit is set to 1 (true); otherwise, it remains 0 (false). This design allows researchers to flexibly combine and apply the flags to construct customized, quality-controlled samples according to their specific scientific objectives. All flags and descriptions are detailed in table \ref{tab:flags_sages_bits}.  

\begin{deluxetable*}{rlll}
\tabletypesize{\scriptsize}
\tablewidth{0pt}
\tablecaption{\texttt{FLAGS\_SAGES} bit definitions \label{tab:flags_sages_bits}}
\tablehead{
\colhead{Bit} & \colhead{Mask} & \colhead{Flag} & \colhead{Definition}
}
\startdata
0  & 1     & \texttt{FLAG\_BORDER} &
Detections within 10 pixels of any detector edge or amplifier boundary. \\
1  & 2     & \texttt{FLAG\_SNR} &
Low signal-to-noise ratio reported by Source Extractor (S/N $< 5$). \\
2  & 4     & \texttt{FLAG\_FWHM} &
Abnormally small FWHM (FWHM $< 2$ pixels), unlikely to be a real astrophysical source. \\
3  & 8     & \texttt{FLAG\_ELON} &
Highly elongated sources (ELONGATION $> 5$). \\
4  & 16    & \texttt{FLAG\_SE\_FLAGS} &
Non-zero native Source Extractor\texttt{FLAGS} inherited from the source-extraction stage. \\
5  & 32    & \texttt{FLAG\_GAIA\_DIS} &
Angular separation from the nearest Gaia counterpart exceeds $1.0\arcsec$. \\
6  & 64    & \texttt{FLAG\_FAKE\_SOURCE} &
On/near linear artifacts ,  within 10 pixels (RANSAC). \\
7  & 128   & \texttt{FLAG\_BAD\_MAG} &
Unphysical or failed photometry (magnitude $< 10$ mag or $> 25$ mag). \\
8  & 256   & \texttt{FLAG\_OUTLIERS\_MAG} &
Photometric calibration outlier: residual $> 3\sigma$ relative to the adopted calibration model. \\
9  & 512   & \texttt{FLAG\_LINEAR\_ERR} &
Frame-level issue: outlier fraction ($>3\sigma$) exceeds 10\%. \\
10 & 1024  & \texttt{FLAG\_BRIGHT\_STARS} &
Bright-star contamination (diffraction spikes/halos/blooming; typically $< 12$ mag). \\
\enddata
\tablecomments{\texttt{FLAGS\_SAGES} is a bitmask. A bit is set to 1 if the corresponding condition is met; otherwise it remains 0.}
\end{deluxetable*}

These flags are further propagated as quality indicators throughout subsequent catalog merging, and are retained in the final master catalog for user reference and downstream analyses.

\subsection{Catalog Merging}

Although sources located near the CCD edges may suffer from degraded photometric precision due to optical distortion and imperfect focusing, the substantial overlaps between adjacent fields ensured by the SAGES tiling strategy, sources located away from the image centers are often observed multiple times under independent photometric conditions. In fact, more than half of the detected sources are covered by two or more exposures. By statistically combining these repeated measurements and discarding problematic detections, the overall photometric precision and catalog reliability can be significantly improved. 

To achieve this goal, we developed a robust, uncertainty-aware merging pipeline. Its core principle is to combine repeated measurements in the linear flux domain (rather than in magnitudes) through inverse-variance weighting, preceded by rigorous outlier rejection.

\subsubsection{Error Modeling and Flag Classification}
Before merging, we construct a comprehensive error model for each single-epoch detection. The total magnitude uncertainty 
$\sigma_{m,i}$ is defined as:
\begin{equation}
\sigma_{m,i}^2 = \mathtt{MAGERR\_AUTO}_i^2 + \sigma_{\mathrm{floor}}^2
\end{equation}

where MAGERR\_AUTO is the photometric error reported by Source Extractor. We add a systematic error floor, $\sigma_{\mathrm{floor}}$ =7 mmag. This value comes from the internal precision validation in section \ref{internalValidation}, to prevent the weights of extremely high-S/N sources from being underestimated due to unmodeled instrumental systematics.

We classify the quality flags into two categories to determine the usability of each detection:
\begin{itemize}
\item \textbf{Hard Flags (Rejection):} Detections affected by severe issues are strictly excluded from the merging process. These include \texttt{FLAG\_FAKE\_SOURCE}, \texttt{FLAG\_BAD\_MAG}, \texttt{FLAG\_FWHM}, \texttt{FLAG\_SNR}, and \texttt{FLAG\_SE\_FLAGS}.
\item \textbf{Soft Flags (Problematic):} Detections with minor issues, such as \texttt{FLAG\_BORDER} or \texttt{FLAG\_ELON}, are retained as fallback options but are prioritized lower than clean detections.
\end{itemize}

All valid magnitudes and their uncertainties are converted into linear flux space:
\begin{equation}
F_i = 10^{-0.4 m_i}, \quad \sigma_{F,i} = \frac{\ln 10}{2.5} F_i \sigma_{m,i}, \quad w_i = \frac{1}{\sigma_{F,i}^2}
\end{equation}

\subsubsection{Merging Strategy}
For detections with a valid Gaia match, grouping is performed based on the unique Gaia source id. Conversely, detections without a Gaia counterpart are treated as distinct, single epoch sources. The pipeline selects the optimal set of measurements based on the following hierarchical decision tree. 

\begin{description}
    \item[\textbf{S ($N_{\rm clean} \ge 2$)}] 
    If two or more "clean" detections exist (FLAGS\_SAGES=0), all are used to compute the inverse-variance weighted average flux. This provides the most robust photometry and error estimation.
    
    \item[\textbf{A ($N_{\rm clean} = 1$)}] 
    If exactly one clean detection is available, its measurement is adopted directly. 
    
    \item[\textbf{B ($N_{\rm clean} = 0, N_{\rm prob} \ge 2$)}] 
    When no clean detections exist but multiple detections with only soft flags (and no hard flags) are available, we do not perform merging. Instead, we adopt the single observation with the highest Signal-to-Noise Ratio (S/N) from this set to minimize the impact of potential artifacts.
    
    \item[\textbf{C ($N_{\rm clean} = 0, N_{\rm prob} = 1$)}] 
    If the source lacks clean detections and possesses exactly one detection with soft flags (and no hard flags), this measurement is adopted.
    
    \item[\textbf{D (Fallback)}] 
    In cases where neither clean nor soft-flagged detections are available ($N_{\rm clean}=0, N_{\rm prob}=0$), but measurements with hard flags exist, we select the single observation with the highest S/N from the remaining group. These measurements are retained as a last resort but are flagged with low quality.
\end{description}
\subsubsection{Flux Combination and Final Error Estimation}
When multiple detections are merged ($N_{\rm used} > 1$), the final flux is the inverse-variance weighted mean:
\begin{equation}
\bar{F} = \frac{\sum w_i F_i}{\sum w_i}
\end{equation}
where $w_i = 1/\sigma_i^2$ is the weight of each observation.

Determining the uncertainty of the merged magnitude is critical. A purely statistical propagation often underestimates the true error if systematic variations (e.g., atmospheric changes) exist between epochs. Conversely, the sample standard deviation can be unreliable for small numbers of observations. We therefore adopt a conservative approach by taking the maximum of the propagated statistical error ($\sigma_{\bar{F},\mathrm{stat}}$) and the standard error of the mean derived from the sample scatter ($\sigma_{\bar{F},\mathrm{scatter}}$).

The propagated statistical error is given by:
\begin{equation}
\sigma_{\bar{F},\mathrm{stat}} = \frac{1}{\sqrt{\sum w_i}}
\end{equation}

For the scatter-based error, we first calculate the unbiased weighted sample variance ($S_F^2$) using the reliability weights:
\begin{equation}
S_F^2 = \frac{V_1}{V_1^2 - V_2} \sum_{i=1}^{N} w_i (F_i - \bar{F})^2
\end{equation}
where $V_1 = \sum w_i$ and $V_2 = \sum w_i^2$. The standard error of the mean based on the observed scatter is then estimated as:
\begin{equation}
\sigma_{\bar{F},\mathrm{scatter}} = \frac{S_F}{\sqrt{N_{\rm used}}}
\end{equation}

The final flux error is adopted as $\sigma_{\bar{F},\mathrm{comb}}=\max(\sigma_{\bar{F},\mathrm{stat}}, \sigma_{\bar{F},\mathrm{scatter}})$. This is converted to magnitude units and added in quadrature with a final systematic error floor ($\sigma_{\mathrm{sys,final}}=7$\,mmag), to account for residual calibration systematics (as motivated by the external validation in Section~\ref{externalValidation}), to obtain the reported error:
\begin{equation}
\sigma_{\bar{m}} = \sqrt{ \left( \frac{2.5}{\ln 10} \frac{\sigma_{\bar{F},\mathrm{comb}}}{\bar{F}} \right)^2 + \sigma_{\mathrm{sys,final}}^2 }
\end{equation}

\subsubsection{Variability and Astrometry}
To identify potential variables or outlier measurements, we compute the $\chi^2$ statistic for the merged group. 
When $N_{\rm used}>1$, we compute internal-consistency diagnostics. For $N_{\rm used}=2$, we use the closed-form
\begin{equation}
\chi^2 = \frac{(F_1-F_2)^2}{\sigma_{F,1}^2+\sigma_{F,2}^2}, \qquad \nu=1,
\end{equation}
and for $n_{\mathrm{used}}>2$,
\begin{equation}
\chi^2 = \sum (F_i-F_{\mathrm{final}})^2 w_i, \qquad \nu=n_{\mathrm{used}}-1.
\end{equation}
Sources with 
\textbf{$\chi^2>3$} and a $ \text{p-value} <0.01$
 are flagged with \texttt{var}.
The final celestial coordinates (RA, Dec) are calculated as the weighted mean of the individual positions. Special care is taken to handle the Right Ascension wrap-around problem for sources near the 0/360
 boundary. The final catalog reports the merged magnitude, the robust uncertainty, the number of observations used ($\text{N}_{used}$), and the detailed quality flags.

\subsection{Post-processing and Flagging}

\subsubsection{Galaxy Identification}

The scientific objective of SAGES focuses on stellar atmospheric parameters; consequently, contamination by extragalactic sources must be minimized. However, Source Extractor's \texttt{MAG\_AUTO} tends to overestimate the brightness of extended sources compared to fixed apertures. As a result, galaxies often appear as outliers in color-color diagrams, potentially biasing stellar parameter estimation. 

To identify these contaminants, we cross-matched our catalog with the REGALADE galaxy catalog \citep{tranin2025catalog}, a comprehensive compilation combining major existing catalogs, deep imaging surveys, and various distance measurements (spectroscopic, photometric, and redshift-independent), contains nearly 80 million galaxies within $D < 2000$ Mpc. First, we applied a fixed matching radius of $1.0\arcsec$ to identify direct matches; in total, 889,250 sources (8\%) were identified and flagged with the field \texttt{galaxy}. Second, to account for source extent and to identify potential contamination from galaxy light, we adopted an adaptive criterion: any source located within $1.5 \times \mathrm{FWHM}$ of a REGALADE galaxy is assigned the \texttt{galaxy\_FWHM} flag. Using this threshold, we flagged 1,106,999 sources (11\%) in our catalog.
\begin{figure}
    \centering
    \includegraphics[width=0.9\linewidth]{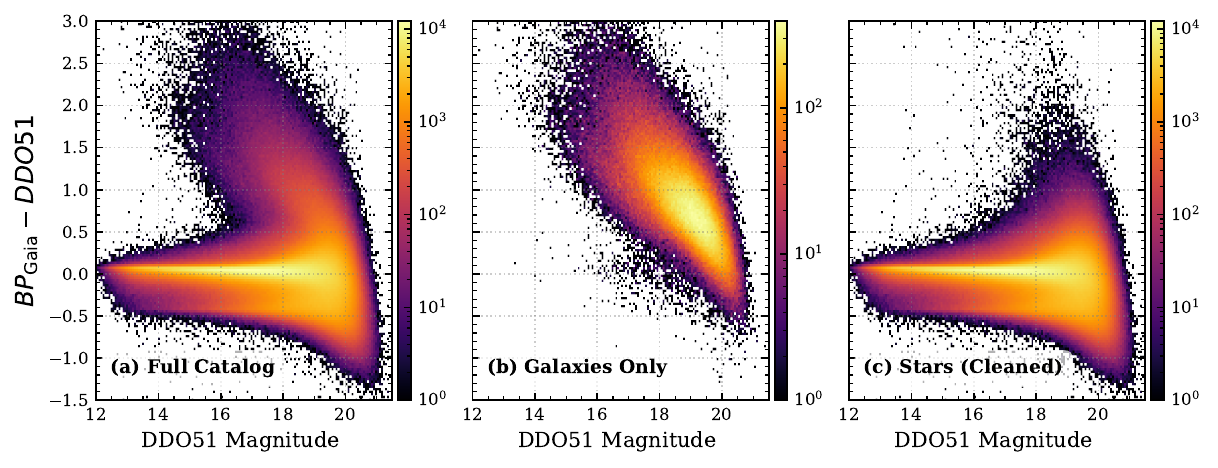}
    \caption{Color–magnitude diagrams in the $(BP_{\mathrm{gaia}} - DDO51)$ versus DDO51 plane. 
    \textbf{(a):} Full sources with good quality and well matched with gaia.
    \textbf{(b):} Subsample of sources matched to the REGALADE galaxy catalog, with \texttt{galaxy=True}  illustrating the locus occupied by galaxies.
    \textbf{(c):} Stellar sample obtained after removing the REGALADE–identified galaxies from the full catalog, yielding a cleaned stellar sequence.
In all panels, the color scale indicates the logarithm of the number density of sources per bin.
    }
    \label{fig:galaxyIdentification}
\end{figure}

To visually validate the classification, Figure~\ref{fig:galaxyIdentification} presents the distribution of sources in the $(BP_{\mathrm{gaia}} - DDO51)$ versus DDO51 plane. The full sources (Figure~\ref{fig:galaxyIdentification}a) exhibit significant scatter outside the main stellar locus. As isolated in (Figure~\ref{fig:galaxyIdentification}b), the identified galaxies occupy a distinct, diffuse region that clearly deviates from the stellar track. By removing these contaminants, we recover a tight and well-defined stellar sequence (Figure~\ref{fig:galaxyIdentification}c), demonstrating the effectiveness of the rejection criteria.
\subsubsection{Non-isolated Source Flagging}

Ground-based observations are intrinsically limited by atmospheric seeing, which can cause multiple astrophysical sources—resolved by space-based missions such as Gaia—to be blended into a single detection. This effect is particularly severe in regions of high stellar density (e.g., near the Galactic plane) and in images obtained under poor seeing conditions, leading to biased or contaminated photometry.

To robustly identify potentially blended sources in the SAGES catalog, we cross-matched each object with the Gaia DR3 catalog and quantified the local source density using three complementary metrics that probe blending on different spatial and astrometric scales:
\begin{description}
	\item[\texttt{n\_gaia\_1p5\_fwhm}] the number of Gaia sources located within a radius of $1.5\times\mathrm{FWHM}$ of the SAGES detection, indicating potential photometric contamination from multiple sources.
	\item[\texttt{n\_gaia\_in\_ellipse}] the number of Gaia sources falling inside the positional uncertainty ellipse of the SAGES source, indicating possible ambiguity in the Gaia cross-identification.
	\item[\texttt{n\_gaia\_10arcsec}] the number of Gaia sources within a fixed radius of ($10\arcsec$), characterizing the local source density and crowding environment.
\end{description}

These three quantities are reported for each source in the catalog, allowing users to apply customized isolation or blending criteria tailored to their specific scientific applications. Rather than adopting a single hard cut, this multi-scale approach preserves flexibility while providing transparent diagnostics of potential flux contamination.

\subsection{Merging Results}
The final merged DDO51 catalog contains a total of 10,489,790 objects, providing a large sample for subsequent analyses.
To ensure robust photometric quality and reliable cross-matching, we define a “clean” high-quality subsample by applying a set of conservative selection criteria. Specifically, we require \texttt{quality\_level} $\in {\texttt{S}, \texttt{A}}$, \texttt{FLAGS\_SAGES} $= 0$, \texttt{var\_flag} $= \texttt{False}$, \texttt{galaxy\_match} $= \texttt{False}$, and \texttt{n\_gaia\_in\_ellipse} $= 1$, thereby excluding sources affected by known data-quality issues, variability, galaxy contamination, or ambiguous Gaia associations. 

Figure~\ref{fig:ddo51_depth} summarizes the photometric depth and precision of the final catalog and the clean subsample.
Panel~(a) shows the DDO51 magnitude distributions for all sources and for the clean subsample, highlighting the effect of the quality cuts on the retained sample size.
Panels~(b) and (c) present the photometric uncertainty as a function of DDO51 magnitude for individual sources in the full catalog and in the clean subsample, respectively. The background density maps are color-coded by the logarithmic number of sources per bin, while the red curves trace the running median uncertainties.
For the clean subsample, the median photometric uncertainty reaches 0.05 and 0.10~mag at fainter magnitudes than in the full sample, with a correspondingly reduced dispersion at fixed magnitude, indicating improved photometric quality relative to the unfiltered catalog.

Because SAGES observations are conducted filter-by-filter sequentially, we provide the \texttt{mean\_mjd} (mean Modified Julian Date) for each merged source in the catalog. This allows users to properly account for stellar variability when cross-matching with external time-domain surveys.
\begin{figure*}
\gridline{
  \fig{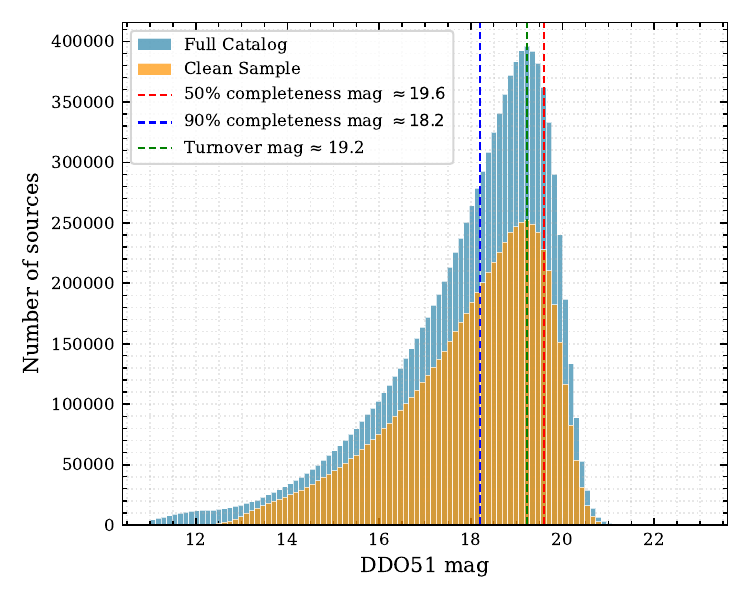}{0.32\textwidth}{(a)}
  \fig{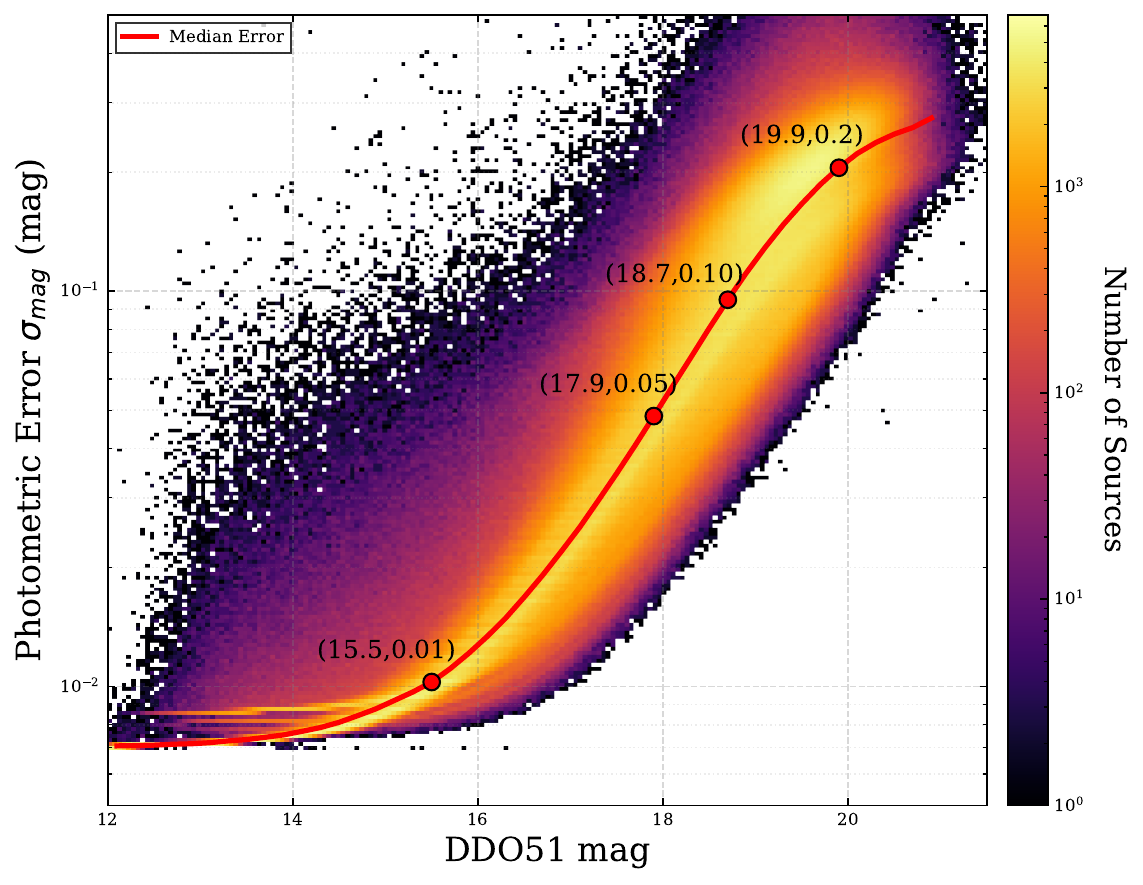}{0.32\textwidth}{(b)}
  \fig{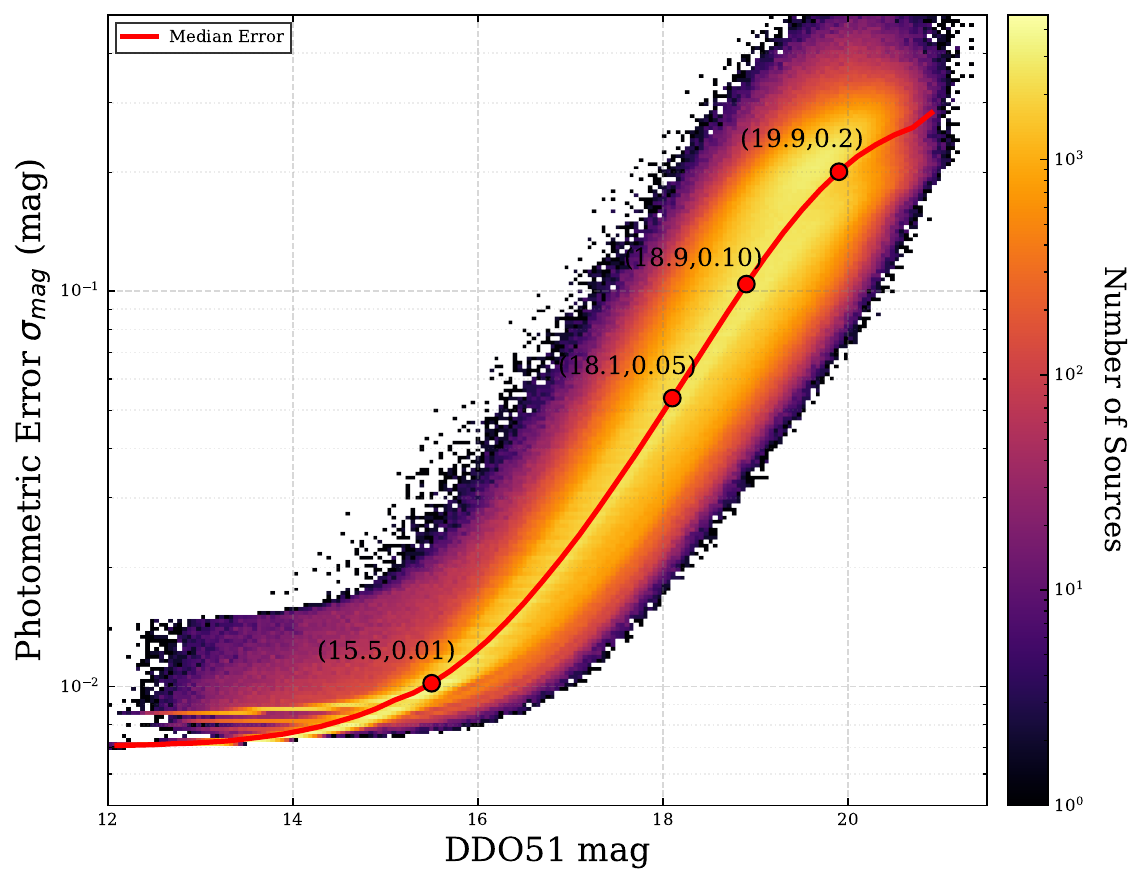}{0.32\textwidth}{(c)}
}
\caption{
DDO51 photometric depth and precision of the merged catalog.  
\textbf{(a)} Histogram of DDO51 magnitudes for all sources and "clean" subsample in the final catalogue, 
showing the number of objects per 0.1-mag bin.  
\textbf{(b)} Photometric uncertainty ($\sigma$) as a function of DDO51 magnitude for individual sources. 
The background is a two-dimensional histogram color–coded by the logarithmic number of sources per bin, 
while the red curve traces the running median $\sigma$. The red circles mark the magnitudes at which 
the median uncertainty reaches $\sim$0.01, 0.05, and 0.10 mag.
\textbf{(c)} Same as panel \textbf{(b)}, but for the "clean" subsample after applying the quality cuts.
}
\label{fig:ddo51_depth}
\end{figure*}
\section{Data Validation}
\label{sec:Validation}
\subsection{Astrometric Validation}

To validate the astrometric accuracy of the final merged catalog, we compared the derived positions of the matched sources with the Gaia DR3 catalog, propagated to the observational epoch of J2023.8. Figure~\ref{fig:final_astrometry} illustrates the distribution of astrometric residuals. The SAGES DDO51 astrometry shows excellent consistency with the Gaia DR3 reference frame. The median positional residual for the entire sample is 87 mas. At the bright end ($G \sim 13.5$ mag), the median residual improves to 14.8 mas, while at the faint limit ($G \sim 21.5$ mag), it increases to approximately 440 mas, consistent with photon noise limits. No significant systematic offsets are observed in either the Right Ascension or Declination directions. 
\begin{figure*}
\gridline{
  \fig{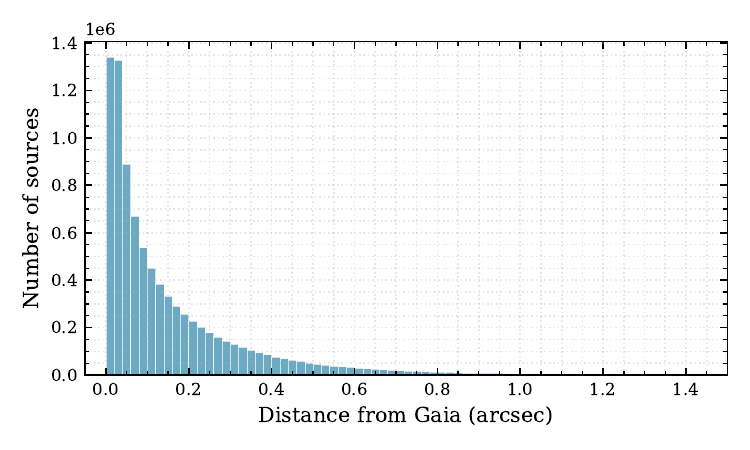}{0.45\textwidth}{(a)}
  \fig{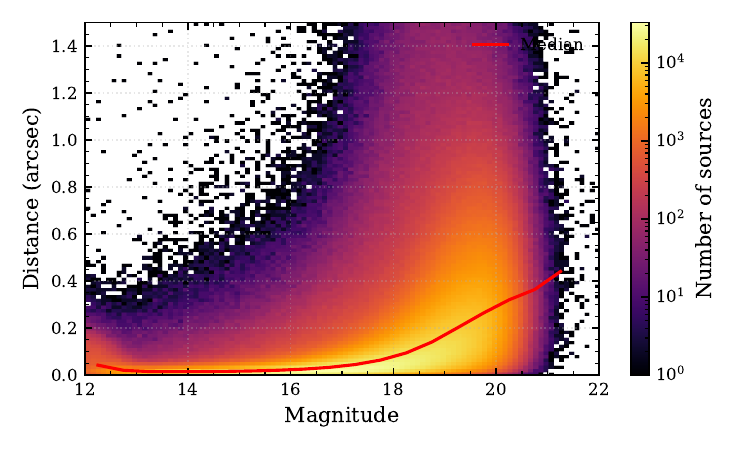}{0.45\textwidth}{(b)}
}

\gridline{
  \fig{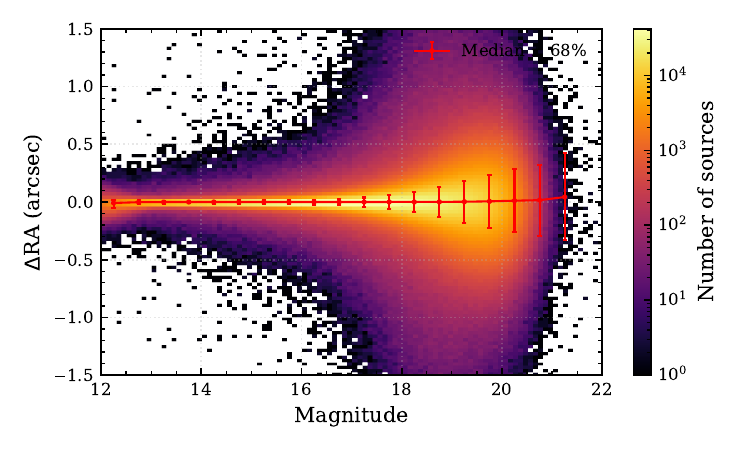}{0.45\textwidth}{(c)}
  \fig{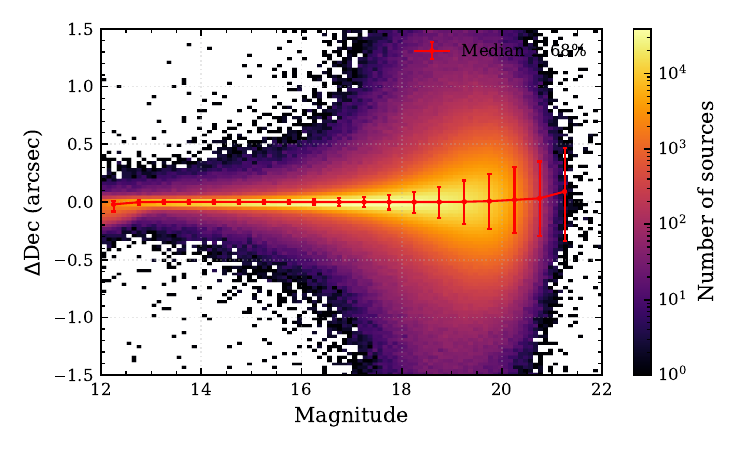}{0.45\textwidth}{(d)}
}

\caption{
Astrometric accuracy of the merged catalog relative to Gaia DR3. 
\textbf{(a)} Histogram of the angular separation between the merged source positions 
and the proper–motion–corrected Gaia coordinates, in arcseconds. 
\textbf{(b)} Angular separation as a function of magnitude; 
the color scale shows the (logarithmic) number of sources per bin and the red curve traces the running median separation. 
\textbf{(c)} Right-ascension residuals ($\Delta$RA) versus magnitude, color–coded by source density; 
the red symbols with error bars indicate the median and the central 68\% interval in each magnitude bin. 
\textbf{(d)} Same as (c), but for declination residuals ($\Delta$Dec).
}
\label{fig:final_astrometry}
\end{figure*}

\subsection{Internal Precision Validation}\label{internalValidation}
To quantify the stability of the calibrated photometry, we utilized both the overlap regions between adjacent fields and specific re-observation campaigns. By selecting a clean sample of point sources (\texttt{FLAGS = 0}), we constructed a dataset of 5.26 million pairs of repeated measurements.

For each pair, we calculated the magnitude difference $\Delta m$ derived from the fully calibrated magnitudes. A robust statistical approach based on the interquartile range was employed to estimate the dispersion ($\sigma_{\Delta m}$) while rejecting outliers. The single-measurement precision was then derived as $\sigma_{\text{int}} = \sigma_{\Delta m} / \sqrt{2}$.

The results are presented in Figure~\ref{fig:internal_precision}. The top panel demonstrates that the median bias (blue dashed line) remains flat and close to zero, indicating a high degree of consistency in our photometric calibration across different exposures. The bottom panel quantifies the precision. We achieve a systematic floor of $\sim 6.6$ mmag for bright stars ($\sim 13.2$ mag). The precision remains better than 1\% (10 mmag) for sources brighter than 15.0 mag, and stays within 10\% down to $\sim$18.5 mag.
\begin{figure}[htbp]
    \centering
    \includegraphics[width=0.5\linewidth]{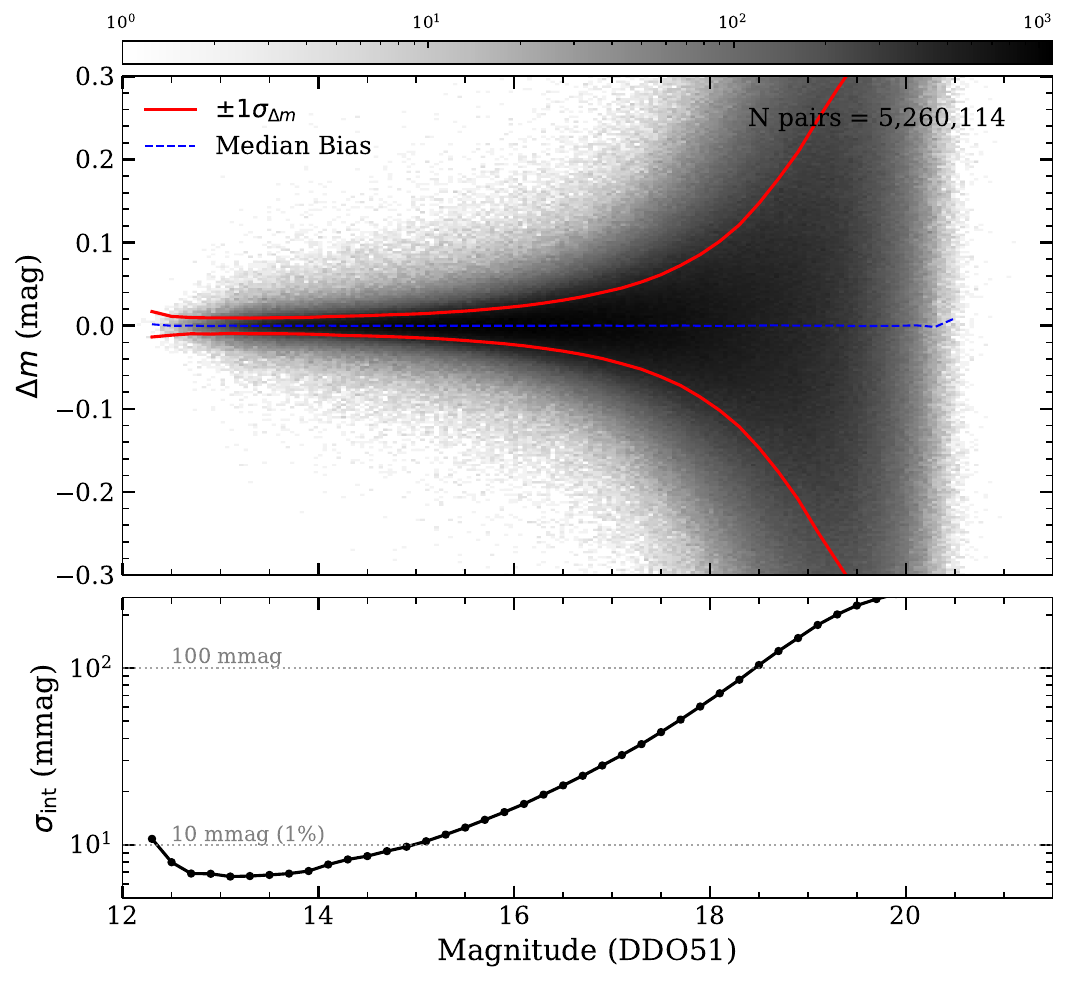} % 
    \caption{Internal photometric precision of SAGES DDO51 derived from 5.26 million pairs of repeated observations. 
    \textbf{Top:} Distribution of magnitude differences ($\Delta m$) for calibrated stars. The red lines mark the $\pm 1\sigma$ envelope, and the blue dashed line shows the median bias.
    \textbf{Bottom:} Single-measurement precision ($\sigma_{\text{int}}$) as a function of magnitude. Dotted lines indicate 1\% (10 mmag) and 10\% precision levels.}
    \label{fig:internal_precision}
\end{figure}

\subsection{External Validation}\label{externalValidation}

To validate the absolute photometric accuracy, we compared our calibrated DDO51 magnitudes of the "clean" subsample with synthetic magnitudes derived from Gaia DR3 XP spectra and the J-PLUS DR3 catalog \citep{cenarro2019j,xiao2023j}, which includes a similar filter $J0515$ with a central wavelength of 5150 \AA\ and an FWHM of 200 \AA. Figure~\ref{fig:external} shows the magnitude residuals for a cross-matched sample of point sources.

The comparison with Gaia DR3 XP synthetic photometry reveals a high degree of consistency at the bright end. The median offset remains flat and essentially zero for stars $<18$ mag, confirming the linearity of our system and the effectiveness of the calibration pipeline in removing instrumental signatures. The scatter ($\sigma$) increases with magnitude as expected from photon statistics. In the bright end, the minimum scatter is about $\sim $8.3 mmag, indicating that our photometric accuracy has reached the sub-percent level.

Comparing  with J-PLUS, we observe a median offset of +16.6 mmag. At the bright end, the minimum scatter is about $\sim $14.5 mmag. These differences are primarily attributed to the different filter transmission profiles. Specifically, the SAGES DDO51 filter is significantly narrower than the J-PLUS $J0515$ band. The narrower bandpass increases the relative weight of the Mg absorption features with respect to the surrounding continuum, thereby yielding systematically fainter magnitudes and producing positive residuals in the comparison between SAGES and J-PLUS.

\begin{figure*}
\gridline{
  \fig{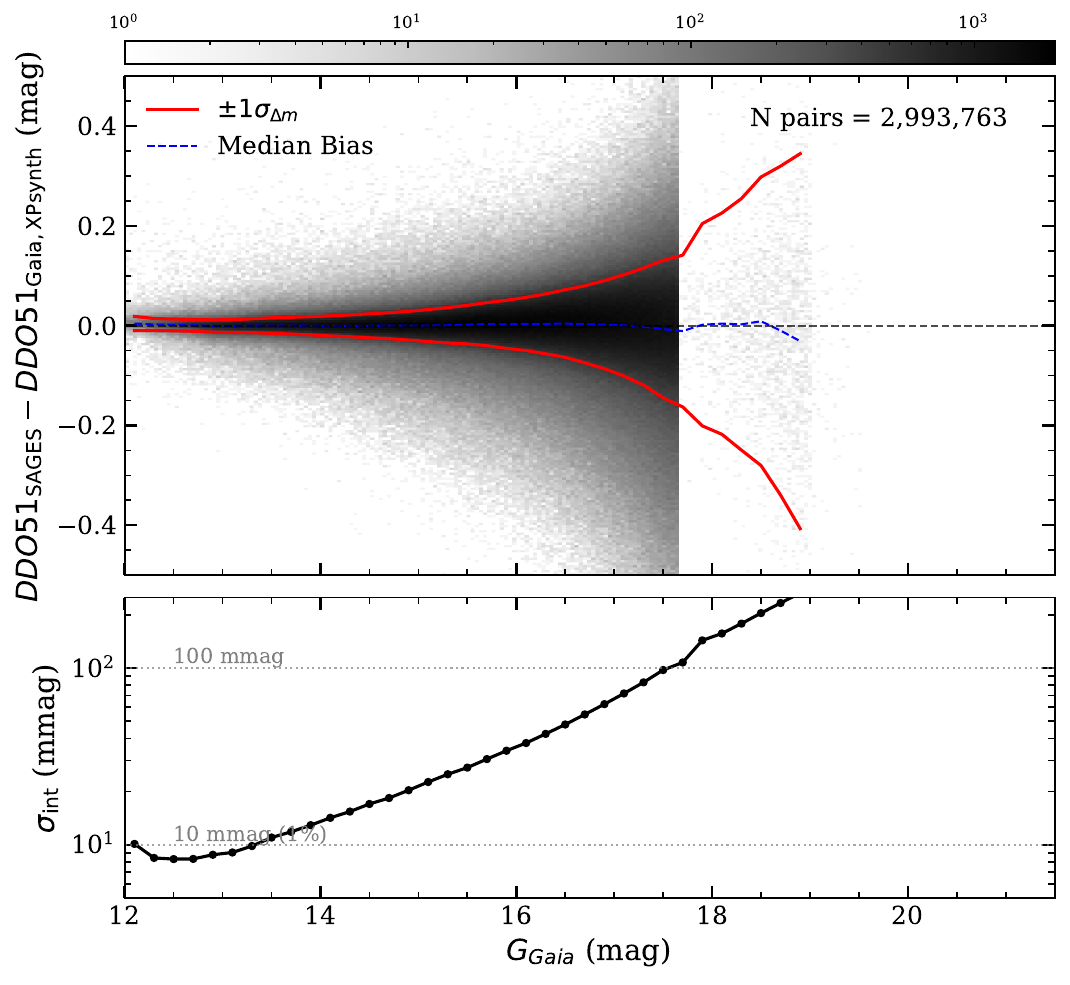}{0.45\textwidth}{\textbf{(a)} SAGES v.s. Gaia XPSP}
  \fig{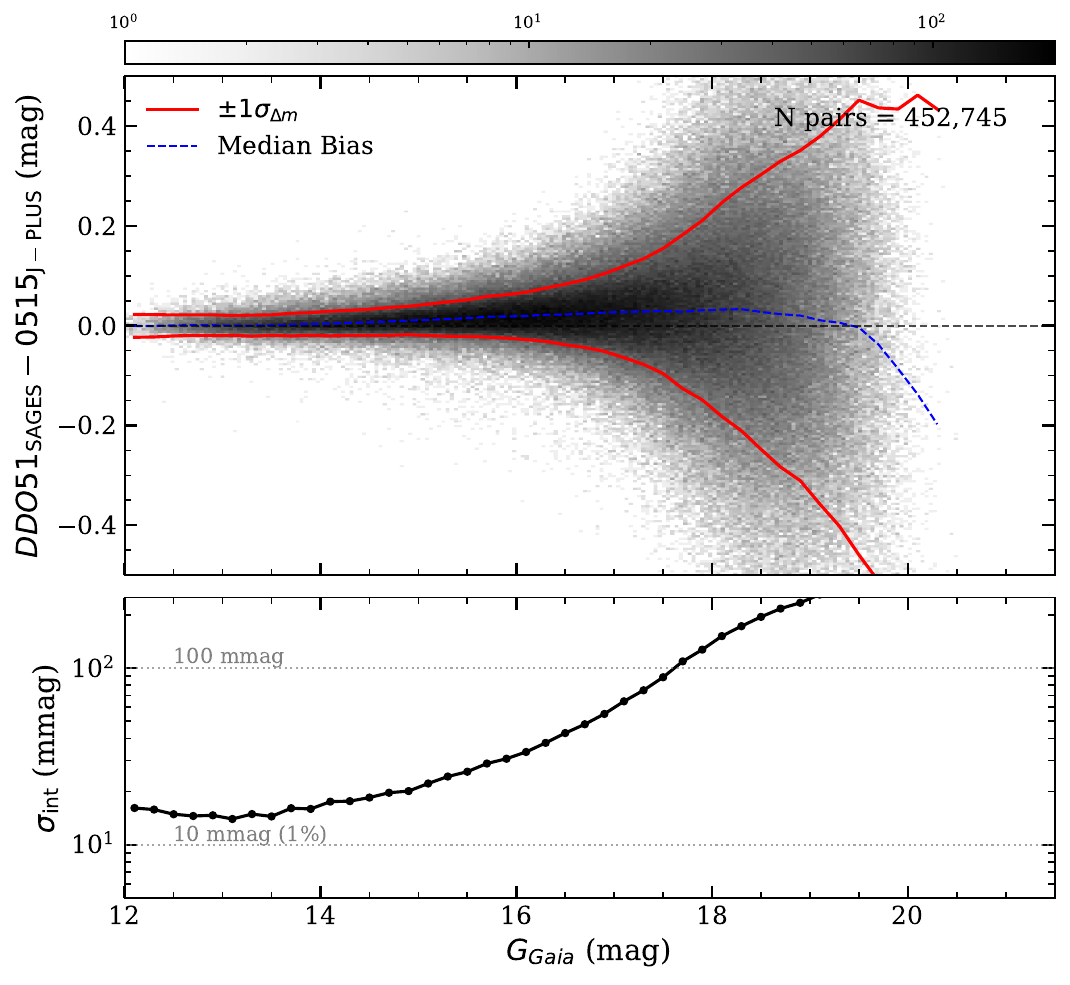}{0.45\textwidth}{\textbf{(b)}SAGES v.s. J-PLUS}
}
\caption{External photometric validation of the SAGES DDO51 calibration. Symbols and color coding follow those in Figure~\ref{fig:internal_precision}.
The panels show the magnitude residuals ($\Delta m = \mathrm{DDO51}_{\mathrm{SAGES}} - m_{\mathrm{ref}}$) as a function of Gaia G magnitude for a cross-matched sample of high-quality point sources. 
\textbf{(a)} Comparison with synthetic DDO51 magnitudes from Gaia XP spectra. \textbf{(b)} Comparison with J-PLUS $J0515$ photometry.
}
\label{fig:external}
\end{figure*}

\section{Scientific demonstration} 

\label{sec:logg}

Using the "clean" subsample described above, we combined our catalog with Gaia DR3 photometry to examine the stellar loci. Figure~\ref{fig:dwarf_giant} presents the observed (not extinction-corrected) color-color diagram of $(G_{BP} - \text{DDO51})$ versus $(G_{BP} - G_{RP})$. The $(G_{BP} - G_{RP})$ index indicates effective temperature, while $(G_{BP} - \mathrm{DDO51})$ traces surface gravity features.

As illustrated in the left panel (Figure~\ref{fig:dwarf_giant}a), a clear bifurcation emerges for cool stars with $(G_{BP} - G_{RP})> 1.0$ ($T_{\rm eff} \lesssim 5500\,{\rm K}$) The upper sequence corresponds to the giant branch, while the lower sequence traces the dwarf branch. To quantify this separation, the right panel (Figure~\ref{fig:dwarf_giant}b) shows the distribution of the gravity-sensitive color index as a function of magnitude for a selected color slice of 
$1.65 < (G_{BP} - G_{RP}) < 1.75$. The two populations display a distinct separation of approximately $\sim 0.3$ mag. While the dispersion naturally increases at fainter magnitudes due to photometric noise, the bimodal structure remains clearly distinguishable.

It is important to note that this clear separation is achieved using only observed colors. Applying corrections for interstellar extinction would reduce the scatter caused by differential reddening, thereby making the distinction between the giant and dwarf branches even more pronounced.
\begin{figure*}
\gridline{
  \fig{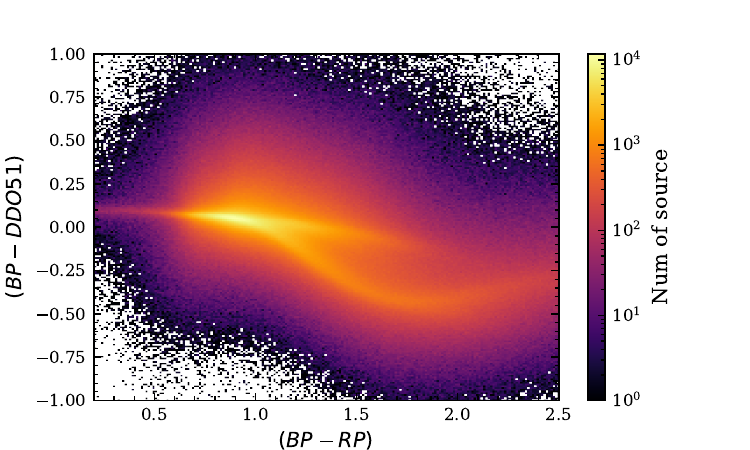}{0.45\textwidth}{(a) Color–Color Diagram}
  \fig{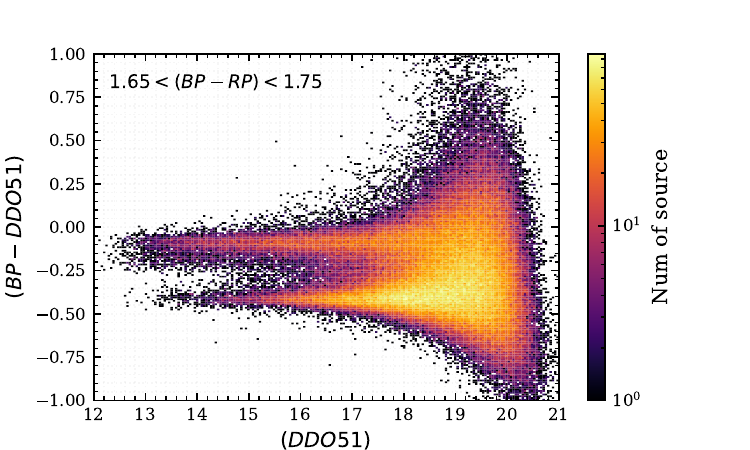}{0.45\textwidth}{(b) Color–Magnitude Diagram}
}
\caption{
Performance of SAGES DDO51 photometry in separating dwarfs and giants.
The sample is using "clean" filtered to exclude galaxies and non-isolated sources.
\textbf{(a)} Distribution of sources in the observed (uncorrected for extinction) 
$(G_{BP} - \text{DDO51})$ versus $(G_{BP} - G_{RP})$ plane. 
The color scale represents the logarithmic number density.  
\textbf{(b)} Variation of the color index $(G_{BP} - \text{DDO51})$ 
with DDO51 magnitude for stars in the range $1.65 < (G_{BP} - G_{RP}) < 1.75$. 
}
\label{fig:dwarf_giant}
\end{figure*}

\section{Data Release and Access}
\label{sec:access}
We provide  a master catalog, details are in Table \ref{tab:datamodel}, which includes 10,489,790 objects. This dataset will be available to the community through the  China-VO platform\footnote{ https://doi.org/10.12149/101721}.
\begin{deluxetable*}{llcl}
\tablecaption{Description of the SAGES DDO51 Photometric Catalog Columns.\label{tab:datamodel}}
\tablehead{
\colhead{Column Name} & \colhead{Format} & \colhead{Unit} & \colhead{Description}
}
\startdata
\texttt{id} & Long & --- & Unique internal identifier for the source \\
\texttt{ra} & Double & deg & Right Ascension (ICRS, J2023.8), weighted mean \\
\texttt{dec} & Double & deg & Declination (ICRS, J2023.8), weighted mean \\
\texttt{ra\_err} & Float & mas & Astrometric uncertainty in RA  \\
\texttt{dec\_err} & Float & mas & Astrometric uncertainty in Dec \\
\texttt{gaia\_source\_id} & Long & --- & Matched Gaia DR3 source identifier \\
\texttt{mag} & Float & mag & Calibrated DDO51 magnitude \\
\texttt{mag\_err} & Float & mag & Total photometric uncertainty (statistical + systematic floor) \\
\texttt{mean\_snr} & Float & --- & Mean Signal-to-Noise Ratio of component exposures \\
\texttt{mean\_fwhm} & Float & arcsec & Mean FWHM of the source \\
mean\_mjd & Float & --- & Mean Modified Julian Date of the combined exposures  \\
\texttt{n\_total} & Short & --- & Total number of available exposures \\
\texttt{n\_used} & Short & --- & Number of exposures used for the merged solution \\
\texttt{var} & Boolean & --- & Variability flag (True if $p < 0.01$ and $\chi^2_{\nu} > 3$) \\
\texttt{merge\_quality} & String & --- & Merging quality grade (S/A/B/C/D) \\
\texttt{chi2\_dof} & Float & --- & Reduced Chi-square ($\chi^2_{\nu}$) of the flux measurement \\
\texttt{p\_value} & Float & --- & Probability of the observed $\chi^2$ (null hypothesis: constant flux) \\
\texttt{FLAGS\_SAGES} & Long & --- & Internal quality bitmask (0 = Clean) \\
\texttt{FLAGS\_SE} & Long & --- & Source Extractor flags from the reference image \\
\texttt{n\_gaia\_1p5\_fwhm} & Short & --- & Number of Gaia sources within $1.5\times$FWHM (blending indicator) \\
\texttt{n\_gaia\_in\_ellipse} & Short & --- & Number of Gaia sources within the positional error ellipse \\
\texttt{n\_gaia\_10arcsec} & Short & --- & Number of Gaia neighbors within $10\arcsec$ (isolation indicator) \\
\texttt{galaxy\_match} & Boolean & --- & Flag: Source is matched to a known galaxy within $1\arcsec$ \\
\texttt{galaxy\_blend} & Boolean & --- & Flag: Source is within $1.5\times$FWHM of a known galaxy \\
\texttt{nearest\_galaxy\_dis} & Float & arcsec & Angular separation to the nearest known galaxy \\
\enddata

\end{deluxetable*}

\section{Summary and Future Work}
\label{sec:summary}
In this paper, we present the first public data release of the SAGES DDO51 band. This release covers a footprint of approximately 2,500 deg$^2$ of the northern sky, comprising over 10 million sources. By implementing a novel calibration strategy anchored to Gaia DR3 XP synthetic photometry, we have achieved millimagnitude-level internal precision and high external accuracy across the survey area.

The DDO51 filter, centered on the \ion{Mg}{1}~$b$ triplet, provides a powerful diagnostic for luminosity classification. As shown in our validation analysis, the combination of SAGES DDO51 photometry with other photometric data, allows for a clear separation between dwarf and giant stars, even without de-reddening. Furthermore, when integrated with the full SAGES multi-band system, this dataset enables precise estimation of stellar surface gravity and metallicities for millions of stars.

Observations for the SAGES DDO51 component are currently ongoing, and observation of H$\alpha_{\rm w}$ will start soon. We aim to complete the survey in the coming observing seasons. Future data releases will not only expand the sky coverage but also include the H$\alpha_{\rm w}$ and H$\alpha_{\rm n}$ band data, further enhancing survey's ability to map interstellar extinction and determine stellar atmospheric parameters. The final, complete SAGES catalog will serve as a legacy resource for Galactic archaeology in the era of large-scale photometric surveys.

%% Please use the acknowledgment and contribution environments. This will 
%% be anonomyized when the "anonymous" style option is used. 
\begin{acknowledgments}

This work is supported by the National Natural Science Foundation of China (NSFC) under grant No.12588202, National Key R\&D Program of China No.2023YFE0107800, No.2024YFA1611900. This work is supported by the Strategic Priority Research Program of the Chinese Academy of Sciences, Grant No. XDB0550100.
ZG acknowledges funding under Department of Human Resources and Social Security of Xinjiang Uygur Autonomous Region Introduced Project “Tianchi talent”. WW is supported by the National Natural Science
Foundation of China grants 62127901, the National Astronomical Observatories Chinese Academy of Sciences No. E4TQ2101, the China Manned Space Project with NO. CMS-CSST-2025-A16 and the Pre-research project on Civil Aerospace Technologies No. D010301 funded by China National Space Administration (CNSA).

This work has made use of data from the European Space Agency (ESA) mission
Gaia (\url{https://www.cosmos.esa.int/Gaia}), processed by the  Gaia
Data Processing and Analysis Consortium (DPAC,
\url{https://www.cosmos.esa.int/web/Gaia/dpac/consortium}). Funding for the DPAC
has been provided by National institutions, in particular the institutions
participating in the Gaia Multilateral Agreement.

Data resources are supported by China National Astronomical Data Center (NADC) and Chinese Virtual Observatory (China-VO). This work is supported by Astronomical Big Data Joint Research Center, co-founded by National Astronomical Observatories, Chinese Academy of Sciences and Alibaba Cloud.
\end{acknowledgments}

% \begin{contribution}
% %%This section gives authors the space to recognize author contributions. The text inside this environment is NOT counted towards the total word quanta. At a minimum, manuscripts are expected to include this text:

% All authors contributed equally to the Terra Mater collaboration.

%% But authors are expected to provide more specific details, e.g. 
%%
%%SC was responsible for writing and submitting the manuscript.
%%WWM came up with the initial research concept and edited the manuscript.
%%OTS obtained the funding and edited the manuscript.
%%EBF provided the formal analysis and validation. He also edited the manuscript.
%%GEH Supervised the undergraduates, wrote the software and administers the project github and Zenodo repositories.
%%
%% Authors can use the Contributor Role Taxonomy (CRediT) at
%% for ideas on how write a good statement tailored to their needs.

% \end{contribution}

\facilities{NOWT:1m , Gaia}

\software{astropy \citep{2013A&A...558A..33A,2018AJ....156..123A,2022ApJ...935..167A},  
          Source Extractor \citep{bertin1996sextractor},
          SCAMP \citep{bertin2006automatic},
          Astrometry.net \citep{lang2010astrometry},
          WCSTools \citep{mink2002wcstools},
          PyGaia \citep{Brown_PyGaia_2024}
          }


%% For this sample we use BibTeX plus aasjournalv7.bst to generate the

\bibliography{sample701}{}

@article{fan2023stellar,
  title={The Stellar Abundances and Galactic Evolution Survey (SAGES). I. General Description and the First Data Release (DR1)},
  author={Fan, Zhou and Zhao, Gang and Wang, Wei and Zheng, Jie and Zhao, Jingkun and Li, Chun and Chen, Yuqin and Yuan, Haibo and Li, Haining and Tan, Kefeng and others},
  journal={The Astrophysical Journal Supplement Series},
  volume={268},
  number={1},
  pages={9},
  year={2023},
  publisher={IOP Publishing}
}

@article{shan2021photometry,
  title={Photometry System of the Nanshan One-meter Wide Telescope at Xinjiang Astronomical Observatory},
  author={Shan, Xing-mei and Zhong, Jing and Zhang, Yu and Niu, Hu-biao and Liu, Jin-zhong and Chen, Li and Li, Jing},
  journal={Progress in Astronomy},
  volume={39},
  pages={118--127},
  year={2021}
}

@article{geisler1984luminosity,
  title={Luminosity classification with the Washington system.},
  author={Geisler, DOUGLAS},
  journal={Publications of the Astronomical Society of the Pacific},
  volume={96},
  number={583},
  pages={723},
  year={1984},
  publisher={IOP Publishing}
}

@article{huang2023beyond,
  title={Beyond Spectroscopy. II. Stellar Parameters for over 20 Million Stars in the Northern Sky from SAGES DR1 and Gaia DR3},
  author={Huang, Yang and Beers, Timothy C and Yuan, Haibo and Tan, Ke-Feng and Wang, Wei and Zheng, Jie and Li, Chun and Lee, Young Sun and Li, Hai-Ning and Zhao, Jing-Kun and others},
  journal={The Astrophysical Journal},
  volume={957},
  number={2},
  pages={65},
  year={2023},
  publisher={IOP Publishing}
}

@article{bessell2005standard,
  title={Standard photometric systems},
  author={Bessell, Michael S},
  journal={Annu. Rev. Astron. Astrophys.},
  volume={43},
  number={1},
  pages={293--336},
  year={2005},
  publisher={Annual Reviews}
}

@article{york2000sloan,
  title={The sloan digital sky survey: Technical summary},
  author={York, Donald G and Adelman, J and Anderson Jr, John E and Anderson, Scott F and Annis, James and Bahcall, Neta A and Bakken, JA and Barkhouser, Robert and Bastian, Steven and Berman, Eileen and others},
  journal={The Astronomical Journal},
  volume={120},
  number={3},
  pages={1579},
  year={2000},
  publisher={IOP Publishing}
}

@article{casagrande2014synthetic,
  title={Synthetic stellar photometry--I. General considerations and new transformations for broad-band systems},
  author={Casagrande, Luca and VandenBerg, Don A},
  journal={Monthly Notices of the Royal Astronomical Society},
  volume={444},
  number={1},
  pages={392--419},
  year={2014},
  publisher={The Royal Astronomical Society}
}

@article{ivezic2008milky,
  title={The milky way tomography with sdss. ii. stellar metallicity},
  author={Ivezi{\'c}, {\v{Z}}eljko and Sesar, Branimir and Juri{\'c}, Mario and Bond, Nicholas and Dalcanton, Julianne and Rockosi, Constance M and Yanny, Brian and Newberg, Heidi J and Beers, Timothy C and Prieto, Carlos Allende and others},
  journal={The Astrophysical Journal},
  volume={684},
  number={1},
  pages={287},
  year={2008},
  publisher={IOP Publishing}
}

@article{cenarro2019j,
  title={J-PLUS: The javalambre photometric local universe survey},
  author={Cenarro, AJ ea and Moles, M and Crist{\'o}bal-Hornillos, David and Mar{\'\i}n-Franch, A and Ederoclite, A and Varela, J and L{\'o}pez-Sanjuan, C and Hern{\'a}ndez-Monteagudo, C and Angulo, RE and Rami{\'o}, H V{\'a}zquez and others},
  journal={Astronomy \& Astrophysics},
  volume={622},
  pages={A176},
  year={2019},
  publisher={EDP Sciences}
}

@inproceedings{kaiser2002pan,
  title={Pan-STARRS: a large synoptic survey telescope array},
  author={Kaiser, Nicholas and Aussel, Herve and Burke, Barry E and Boesgaard, Hans and Chambers, Ken and Chun, Mark Richard and Heasley, James N and Hodapp, Klaus-Werner and Hunt, Bobby and Jedicke, Robert and others},
  booktitle={Survey and Other Telescope Technologies and Discoveries},
  volume={4836},
  pages={154--164},
  year={2002},
  organization={SPIE}
}

@article{prusti2016gaia,
  title={The gaia mission},
  author={Prusti, Timo and De Bruijne, JHJ and Brown, Anthony GA and Vallenari, Antonella and Babusiaux, C and Bailer-Jones, CAL and Bastian, U and Biermann, M and Evans, Dafydd Wyn and Eyer, L and others},
  journal={Astronomy \& astrophysics},
  volume={595},
  pages={A1},
  year={2016},
  publisher={EDP sciences}
}

@article{ivezic2012galactic,
  title={Galactic stellar populations in the era of the sloan digital sky survey and other large surveys},
  author={Ivezi{\'c}, {\v{Z}}eljko and Beers, Timothy C and Juri{\'c}, Mario},
  journal={Annual Review of Astronomy and Astrophysics},
  volume={50},
  number={1},
  pages={251--304},
  year={2012},
  publisher={Annual Reviews}
}

@article{perryman2025space,
  title={Space Astrometry with Gaia: Advances in Understanding our Galaxy},
  author={Perryman, Michael},
  journal={arXiv preprint arXiv:2509.10883},
  year={2025}
}

@article{gu2025stellar,
  title={The Stellar Abundances and Galactic Evolution Survey (SAGES). II. Machine Learning--based Stellar Parameters for 21 Million Stars from the First Data Release},
  author={Gu, Hongrui and Fan, Zhou and Zhao, Gang and Yang, Huang and Beers, Timothy C and Wang, Wei and Zheng, Jie and Zhao, Jingkun and Li, Chun and Chen, Yuqin and others},
  journal={The Astrophysical Journal Supplement Series},
  volume={277},
  number={1},
  pages={19},
  year={2025},
  publisher={IOP Publishing}
}

@article{hong2024candidate,
  title={Candidate Members of the VMP/EMP Disk System of the Galaxy from the SkyMapper and SAGES Surveys},
  author={Hong, Jihye and Beers, Timothy C and Lee, Young Sun and Huang, Yang and Hirai, Yutaka and Garcia, Jonathan Cabrera and Shank, Derek and Xu, Shuai and Yuan, Haibo and Mardini, Mohammad K and others},
  journal={The Astrophysical Journal Supplement Series},
  volume={273},
  number={1},
  pages={12},
  year={2024},
  publisher={IOP Publishing}
}

@article{li2024stellar,
  title={The Stellar Abundances and Galactic Evolution Survey (SAGES) III--The g/r/i-band Data Release},
  author={Li, Chun and Fan, Zhou and Zhao, Gang and Wang, Wei and Zheng, Jie and Tan, Kefeng and Zhao, Jingkun and Huang, Yang and Yuan, Haibo and Xiao, Kai and others},
  journal={arXiv preprint arXiv:2410.10218},
  year={2024}
}

@article{zheng2024strategies,
  title={The Strategies and Scheduler Program for the SAGES Sky Survey},
  author={ZHENG, Jie and WANG, Wei and FAN, Zhou and LI, Chun and ZHAO, Gang},
  journal={Progress in Astronomy},
  volume={42},
  pages={698--708},
  year={2024}
}

@article{bai2020wide,
  title={The wide-field photometric system of the Nanshan One-meter Telescope},
  author={Bai, Chun-Hai and Feng, Guo-Jie and Zhang, Xuan and Niu, Hu-Biao and Eskandar, Abdusamatjan and Pu, Guang-Xin and Ma, Shu-Guo and Liu, Jin-Zhong and Jiang, Xiao-Jun and Ma, Lu and others},
  journal={Research in Astronomy and Astrophysics},
  volume={20},
  number={12},
  pages={211},
  year={2020},
  publisher={IOP Publishing}
}

@article{yuan2015stellar,
  title={Stellar color regression: a spectroscopy-based method for color calibration to a few millimagnitude accuracy and the recalibration of Stripe 82},
  author={Yuan, Haibo and Liu, Xiaowei and Xiang, Maosheng and Huang, Yang and Zhang, Huihua and Chen, Bingqiu},
  journal={The Astrophysical Journal},
  volume={799},
  number={2},
  pages={133},
  year={2015},
  publisher={IOP Publishing}
}

@article{xiao2023,
  author={Xiao, Kai and Yuan, Haibo and Huang, Bowen and Xu, Shuai and Zheng, Jie and Li, Chun and Fan, Zhou and Wang, Wei and Zhao, Gang and Feng, Guojie and others},
  title = "Photometric calibration of the Stellar Abundance and Galactic Evolution Survey (SAGES): Nanshan One-meter Wide-field Telescope g, r, and i band imaging data",
  journal = "Chinese Science Bulletin",
  year = "2023",
  volume = "68",
  number = "21",
  pages = "2790-2804",
}

@article{bertin1996sextractor,
  title={SExtractor: Software for source extraction},
  author={Bertin, Emmanuel and Arnouts, Stephane},
  journal={Astronomy and astrophysics supplement series},
  volume={117},
  number={2},
  pages={393--404},
  year={1996},
  publisher={EDP Sciences}
}

@inproceedings{bertin2006automatic,
  title={Automatic astrometric and photometric calibration with SCAMP},
  author={Bertin, E},
  booktitle={Astronomical Data Analysis Software and Systems XV},
  volume={351},
  pages={112},
  year={2006}
}

@article{lang2010astrometry,
  title={Astrometry. net: Blind astrometric calibration of arbitrary astronomical images},
  author={Lang, Dustin and Hogg, David W and Mierle, Keir and Blanton, Michael and Roweis, Sam},
  journal={The astronomical journal},
  volume={139},
  number={5},
  pages={1782},
  year={2010},
  publisher={IOP Publishing}
}

@article{tranin2025catalog,
  title={A catalog to unite them all: REGALADE, a revised galaxy compilation for the advanced detector era},
  author={Tranin, Hugo and Blagorodnova, Nadejda and G{\'o}mez-Mu{\~n}oz, Marco A and Wavasseur, Maxime and Groot, Paul J and Landsberg, Lloyd and Stoppa, Fiorenzo and Bloemen, Steven and Vreeswijk, Paul M and Pieterse, Dani{\"e}lle LA and others},
  journal={arXiv preprint arXiv:2508.13267},
  year={2025}
}

@article{wang2013stromgren,
  title={Str{\"o}mgren-Crawford uvby$\beta$ all sky survey-towards understanding of the Galaxy},
  author={Wang, Wei and Zhao, Gang and Chen, Yuqin and Liu, Yujuan},
  journal={Proceedings of the International Astronomical Union},
  volume={9},
  number={S298},
  pages={326--330},
  year={2013},
  publisher={Cambridge University Press}
}

@article{zheng2018sage,
  title={The SAGE photometric survey: technical description},
  author={Zheng, Jie and Zhao, Gang and Wang, Wei and Fan, Zhou and Tan, Ke-Feng and Li, Chun and Zuo, Fang},
  journal={Research in Astronomy and Astrophysics},
  volume={18},
  number={12},
  pages={147},
  year={2018},
  publisher={IOP Publishing}
}

@article{Fan2018sage,
  title={Stellar Abundance and Galaxy Evolution -- Survey Photometric System and Data Reduction},
  author={Fan, Zhou and Zhao, Gang and Wang, Wei and Zhao, jingkun and Zheng, Jie and Tan, kefeng and Liu, Yujuan and He, Wei and Song Yihan and Jiang Xiaojun},
  journal={Progress In Astronomy},
  volume={36},
  number={2},
  pages={101--121},
  year={2018},
  publisher={IOP Publishing}
}

@article{zheng2019sage,
  title={Research on the Data Reduction of the SAGE Photometric Survey},
  author={Zheng, Jie and Zhao, Gang and Wang, Wei and Fan, Zhou and Zhao, jingkun and Tan, kefeng},
  year={2019},
  volume={16},
  number={1},
  pages={93--106},
  journal={Astronomical Research and Technology}
}

@article{zhang2025stellar,
  title={The Stellar Abundances and Galactic Evolution Survey (SAGES). IV. Surface Gravity Estimation and Giant--Dwarf Separation with the DDO51 Filter},
  author={Zhang, Qiqian and Fan, Zhou and Zhao, Gang and Wu, Ying and Wang, Wei and Xiao, Kai and Gu, Hongrui and Zheng, Jie and Zhao, Jingkun and Li, Chun and others},
  journal={The Astrophysical Journal},
  volume={993},
  number={2},
  pages={170},
  year={2025},
  publisher={IOP Publishing}
}

@ARTICLE{2022ApJ...935..167A,
       author = {{Astropy Collaboration} and {Price-Whelan}, Adrian M. and {Lim}, Pey Lian and {Earl}, Nicholas and {Starkman}, Nathaniel and {Bradley}, Larry and {Shupe}, David L. and {Patil}, Aarya A. and {Corrales}, Lia and {Brasseur}, C.~E. and {N{\"o}the}, Maximilian and {Donath}, Axel and {Tollerud}, Erik and {Morris}, Brett M. and {Ginsburg}, Adam and {Vaher}, Eero and {Weaver}, Benjamin A. and {Tocknell}, James and {Jamieson}, William and {van Kerkwijk}, Marten H. and {Robitaille}, Thomas P. and {Merry}, Bruce and {Bachetti}, Matteo and {G{\"u}nther}, H. Moritz and {Aldcroft}, Thomas L. and {Alvarado-Montes}, Jaime A. and {Archibald}, Anne M. and {B{\'o}di}, Attila and {Bapat}, Shreyas and {Barentsen}, Geert and {Baz{\'a}n}, Juanjo and {Biswas}, Manish and {Boquien}, M{\'e}d{\'e}ric and {Burke}, D.~J. and {Cara}, Daria and {Cara}, Mihai and {Conroy}, Kyle E. and {Conseil}, Simon and {Craig}, Matthew W. and {Cross}, Robert M. and {Cruz}, Kelle L. and {D'Eugenio}, Francesco and {Dencheva}, Nadia and {Devillepoix}, Hadrien A.~R. and {Dietrich}, J{\"o}rg P. and {Eigenbrot}, Arthur Davis and {Erben}, Thomas and {Ferreira}, Leonardo and {Foreman-Mackey}, Daniel and {Fox}, Ryan and {Freij}, Nabil and {Garg}, Suyog and {Geda}, Robel and {Glattly}, Lauren and {Gondhalekar}, Yash and {Gordon}, Karl D. and {Grant}, David and {Greenfield}, Perry and {Groener}, Austen M. and {Guest}, Steve and {Gurovich}, Sebastian and {Handberg}, Rasmus and {Hart}, Akeem and {Hatfield-Dodds}, Zac and {Homeier}, Derek and {Hosseinzadeh}, Griffin and {Jenness}, Tim and {Jones}, Craig K. and {Joseph}, Prajwel and {Kalmbach}, J. Bryce and {Karamehmetoglu}, Emir and {Ka{\l}uszy{\'n}ski}, Miko{\l}aj and {Kelley}, Michael S.~P. and {Kern}, Nicholas and {Kerzendorf}, Wolfgang E. and {Koch}, Eric W. and {Kulumani}, Shankar and {Lee}, Antony and {Ly}, Chun and {Ma}, Zhiyuan and {MacBride}, Conor and {Maljaars}, Jakob M. and {Muna}, Demitri and {Murphy}, N.~A. and {Norman}, Henrik and {O'Steen}, Richard and {Oman}, Kyle A. and {Pacifici}, Camilla and {Pascual}, Sergio and {Pascual-Granado}, J. and {Patil}, Rohit R. and {Perren}, Gabriel I. and {Pickering}, Timothy E. and {Rastogi}, Tanuj and {Roulston}, Benjamin R. and {Ryan}, Daniel F. and {Rykoff}, Eli S. and {Sabater}, Jose and {Sakurikar}, Parikshit and {Salgado}, Jes{\'u}s and {Sanghi}, Aniket and {Saunders}, Nicholas and {Savchenko}, Volodymyr and {Schwardt}, Ludwig and {Seifert-Eckert}, Michael and {Shih}, Albert Y. and {Jain}, Anany Shrey and {Shukla}, Gyanendra and {Sick}, Jonathan and {Simpson}, Chris and {Singanamalla}, Sudheesh and {Singer}, Leo P. and {Singhal}, Jaladh and {Sinha}, Manodeep and {Sip{\H{o}}cz}, Brigitta M. and {Spitler}, Lee R. and {Stansby}, David and {Streicher}, Ole and {{\v{S}}umak}, Jani and {Swinbank}, John D. and {Taranu}, Dan S. and {Tewary}, Nikita and {Tremblay}, Grant R. and {de Val-Borro}, Miguel and {Van Kooten}, Samuel J. and {Vasovi{\'c}}, Zlatan and {Verma}, Shresth and {de Miranda Cardoso}, Jos{\'e} Vin{\'\i}cius and {Williams}, Peter K.~G. and {Wilson}, Tom J. and {Winkel}, Benjamin and {Wood-Vasey}, W.~M. and {Xue}, Rui and {Yoachim}, Peter and {Zhang}, Chen and {Zonca}, Andrea and {Astropy Project Contributors}},
        title = "{The Astropy Project: Sustaining and Growing a Community-oriented Open-source Project and the Latest Major Release (v5.0) of the Core Package}",
      journal = {\apj},
         year = 2022,
        month = aug,
       volume = {935},
       number = {2},
          eid = {167},
        pages = {167},
          doi = {10.3847/1538-4357/ac7c74},
archivePrefix = {arXiv},
       eprint = {2206.14220},
 primaryClass = {astro-ph.IM},
       adsurl = {https://ui.adsabs.harvard.edu/abs/2022ApJ...935..167A}
}

@ARTICLE{2018AJ....156..123A,
       author = {{Astropy Collaboration} and {Price-Whelan}, A.~M. and {Sip{\H{o}}cz}, B.~M. and {G{\"u}nther}, H.~M. and {Lim}, P.~L. and {Crawford}, S.~M. and {Conseil}, S. and {Shupe}, D.~L. and {Craig}, M.~W. and {Dencheva}, N. and {Ginsburg}, A. and {VanderPlas}, J.~T. and {Bradley}, L.~D. and {P{\'e}rez-Su{\'a}rez}, D. and {de Val-Borro}, M. and {Aldcroft}, T.~L. and {Cruz}, K.~L. and {Robitaille}, T.~P. and {Tollerud}, E.~J. and {Ardelean}, C. and {Babej}, T. and {Bach}, Y.~P. and {Bachetti}, M. and {Bakanov}, A.~V. and {Bamford}, S.~P. and {Barentsen}, G. and {Barmby}, P. and {Baumbach}, A. and {Berry}, K.~L. and {Biscani}, F. and {Boquien}, M. and {Bostroem}, K.~A. and {Bouma}, L.~G. and {Brammer}, G.~B. and {Bray}, E.~M. and {Breytenbach}, H. and {Buddelmeijer}, H. and {Burke}, D.~J. and {Calderone}, G. and {Cano Rodr{\'\i}guez}, J.~L. and {Cara}, M. and {Cardoso}, J.~V.~M. and {Cheedella}, S. and {Copin}, Y. and {Corrales}, L. and {Crichton}, D. and {D'Avella}, D. and {Deil}, C. and {Depagne}, {\'E}. and {Dietrich}, J.~P. and {Donath}, A. and {Droettboom}, M. and {Earl}, N. and {Erben}, T. and {Fabbro}, S. and {Ferreira}, L.~A. and {Finethy}, T. and {Fox}, R.~T. and {Garrison}, L.~H. and {Gibbons}, S.~L.~J. and {Goldstein}, D.~A. and {Gommers}, R. and {Greco}, J.~P. and {Greenfield}, P. and {Groener}, A.~M. and {Grollier}, F. and {Hagen}, A. and {Hirst}, P. and {Homeier}, D. and {Horton}, A.~J. and {Hosseinzadeh}, G. and {Hu}, L. and {Hunkeler}, J.~S. and {Ivezi{\'c}}, {\v{Z}}. and {Jain}, A. and {Jenness}, T. and {Kanarek}, G. and {Kendrew}, S. and {Kern}, N.~S. and {Kerzendorf}, W.~E. and {Khvalko}, A. and {King}, J. and {Kirkby}, D. and {Kulkarni}, A.~M. and {Kumar}, A. and {Lee}, A. and {Lenz}, D. and {Littlefair}, S.~P. and {Ma}, Z. and {Macleod}, D.~M. and {Mastropietro}, M. and {McCully}, C. and {Montagnac}, S. and {Morris}, B.~M. and {Mueller}, M. and {Mumford}, S.~J. and {Muna}, D. and {Murphy}, N.~A. and {Nelson}, S. and {Nguyen}, G.~H. and {Ninan}, J.~P. and {N{\"o}the}, M. and {Ogaz}, S. and {Oh}, S. and {Parejko}, J.~K. and {Parley}, N. and {Pascual}, S. and {Patil}, R. and {Patil}, A.~A. and {Plunkett}, A.~L. and {Prochaska}, J.~X. and {Rastogi}, T. and {Reddy Janga}, V. and {Sabater}, J. and {Sakurikar}, P. and {Seifert}, M. and {Sherbert}, L.~E. and {Sherwood-Taylor}, H. and {Shih}, A.~Y. and {Sick}, J. and {Silbiger}, M.~T. and {Singanamalla}, S. and {Singer}, L.~P. and {Sladen}, P.~H. and {Sooley}, K.~A. and {Sornarajah}, S. and {Streicher}, O. and {Teuben}, P. and {Thomas}, S.~W. and {Tremblay}, G.~R. and {Turner}, J.~E.~H. and {Terr{\'o}n}, V. and {van Kerkwijk}, M.~H. and {de la Vega}, A. and {Watkins}, L.~L. and {Weaver}, B.~A. and {Whitmore}, J.~B. and {Woillez}, J. and {Zabalza}, V. and {Astropy Contributors}},
        title = "{The Astropy Project: Building an Open-science Project and Status of the v2.0 Core Package}",
      journal = {\aj},
         year = 2018,
        month = sep,
       volume = {156},
       number = {3},
          eid = {123},
        pages = {123},
          doi = {10.3847/1538-3881/aabc4f},
archivePrefix = {arXiv},
       eprint = {1801.02634},
 primaryClass = {astro-ph.IM},
       adsurl = {https://ui.adsabs.harvard.edu/abs/2018AJ....156..123A}
}

@ARTICLE{2013A&A...558A..33A,
       author = {{Astropy Collaboration} and {Robitaille}, Thomas P. and {Tollerud}, Erik J. and {Greenfield}, Perry and {Droettboom}, Michael and {Bray}, Erik and {Aldcroft}, Tom and {Davis}, Matt and {Ginsburg}, Adam and {Price-Whelan}, Adrian M. and {Kerzendorf}, Wolfgang E. and {Conley}, Alexander and {Crighton}, Neil and {Barbary}, Kyle and {Muna}, Demitri and {Ferguson}, Henry and {Grollier}, Fr{\'e}d{\'e}ric and {Parikh}, Madhura M. and {Nair}, Prasanth H. and {Unther}, Hans M. and {Deil}, Christoph and {Woillez}, Julien and {Conseil}, Simon and {Kramer}, Roban and {Turner}, James E.~H. and {Singer}, Leo and {Fox}, Ryan and {Weaver}, Benjamin A. and {Zabalza}, Victor and {Edwards}, Zachary I. and {Azalee Bostroem}, K. and {Burke}, D.~J. and {Casey}, Andrew R. and {Crawford}, Steven M. and {Dencheva}, Nadia and {Ely}, Justin and {Jenness}, Tim and {Labrie}, Kathleen and {Lim}, Pey Lian and {Pierfederici}, Francesco and {Pontzen}, Andrew and {Ptak}, Andy and {Refsdal}, Brian and {Servillat}, Mathieu and {Streicher}, Ole},
        title = "{Astropy: A community Python package for astronomy}",
      journal = {\aap},
         year = 2013,
        month = oct,
       volume = {558},
          eid = {A33},
        pages = {A33},
          doi = {10.1051/0004-6361/201322068},
archivePrefix = {arXiv},
       eprint = {1307.6212},
 primaryClass = {astro-ph.IM},
       adsurl = {https://ui.adsabs.harvard.edu/abs/2013A&A...558A..33A}
}

@inproceedings{mink2002wcstools,
  title={WCSTools 3.0: More tools for image astrometry and catalog searching},
  author={Mink, Douglas J},
  booktitle={Astronomical Data Analysis Software and Systems XI},
  volume={281},
  pages={169},
  year={2002}
}

@article{huang2022photometric,
  title={Photometric calibration methods for wide-field photometric surveys},
  author={Huang, Bowen and Xiao, Kai and Yuan, Haibo},
  journal={SCIENTIA SINICA Physica, Mechanica \& Astronomica},
  year={2022},
  volume = "52",
  number = "8",
  pages = "289503-",
}

@article{huang2024comprehensive,
  title={A comprehensive correction of the Gaia DR3 XP spectra},
  author={Huang, Bowen and Yuan, Haibo and Xiang, Maosheng and Huang, Yang and Xiao, Kai and Xu, Shuai and Zhang, Ruoyi and Yang, Lin and Niu, Zexi and Gu, Hongrui},
  journal={The Astrophysical Journal Supplement Series},
  volume={271},
  number={1},
  pages={13},
  year={2024},
  publisher={IOP Publishing}
}

@article{chambers2016pan,
  title={The pan-starrs1 surveys},
  author={Chambers, Kenneth C and Magnier, EA and Metcalfe, N and Flewelling, HA and Huber, ME and Waters, CZ and Denneau, L and Draper, PW and Farrow, D and Finkbeiner, DP and others},
  journal={arXiv preprint arXiv:1612.05560},
  year={2016}
}

@article{xiao2023j,
  title={J-PLUS: Photometric Recalibration with the Stellar Color Regression Method and an Improved Gaia XP Synthetic Photometry Method},
  author={Xiao, Kai and Yuan, Haibo and Lopez-Sanjuan, C and Huang, Yang and Huang, Bowen and Beers, Timothy C and Xu, Shuai and Wang, Yuanchang and Yang, Lin and Alcaniz, Jailson and others},
  journal={The Astrophysical Journal Supplement Series},
  volume={269},
  number={2},
  pages={58},
  year={2023},
  publisher={IOP Publishing}
}

@article{fischler1981random,
  title={Random sample consensus: a paradigm for model fitting with applications to image analysis and automated cartography},
  author={Fischler, Martin A and Bolles, Robert C},
  journal={Communications of the ACM},
  volume={24},
  number={6},
  pages={381--395},
  year={1981},
  publisher={ACM New York, NY, USA}
}

@article{roser2008ppm,
  title={PPM-Extended (PPMX)--a catalogue of positions and proper motions},
  author={R{\"o}ser, S and Schilbach, E and Schwan, H and Kharchenko, NV and Piskunov, AE and Scholz, R-D},
  journal={Astronomy \& Astrophysics},
  volume={488},
  number={1},
  pages={401--408},
  year={2008},
  publisher={EDP Sciences}
}

@ARTICLE{2023A&A...674A...1G,
       author = {{Gaia Collaboration} and {Vallenari}, A. and {Brown}, A.~G.~A. and {Prusti}, T. and {de Bruijne}, J.~H.~J. and {Arenou}, F. and {Babusiaux}, C. and {Biermann}, M. and {Creevey}, O.~L. and {Ducourant}, C. and {Evans}, D.~W. and {Eyer}, L. and {Guerra}, R. and {Hutton}, A. and {Jordi}, C. and {Klioner}, S.~A. and {Lammers}, U.~L. and {Lindegren}, L. and {Luri}, X. and {Mignard}, F. and {Panem}, C. and {Pourbaix}, D. and {Randich}, S. and {Sartoretti}, P. and {Soubiran}, C. and {Tanga}, P. and {Walton}, N.~A. and {Bailer-Jones}, C.~A.~L. and {Bastian}, U. and {Drimmel}, R. and {Jansen}, F. and {Katz}, D. and {Lattanzi}, M.~G. and {van Leeuwen}, F. and {Bakker}, J. and {Cacciari}, C. and {Casta{\~n}eda}, J. and {De Angeli}, F. and {Fabricius}, C. and {Fouesneau}, M. and {Fr{\'e}mat}, Y. and {Galluccio}, L. and {Guerrier}, A. and {Heiter}, U. and {Masana}, E. and {Messineo}, R. and {Mowlavi}, N. and {Nicolas}, C. and {Nienartowicz}, K. and {Pailler}, F. and {Panuzzo}, P. and {Riclet}, F. and {Roux}, W. and {Seabroke}, G.~M. and {Sordo}, R. and {Th{\'e}venin}, F. and {Gracia-Abril}, G. and {Portell}, J. and {Teyssier}, D. and {Altmann}, M. and {Andrae}, R. and {Audard}, M. and {Bellas-Velidis}, I. and {Benson}, K. and {Berthier}, J. and {Blomme}, R. and {Burgess}, P.~W. and {Busonero}, D. and {Busso}, G. and {C{\'a}novas}, H. and {Carry}, B. and {Cellino}, A. and {Cheek}, N. and {Clementini}, G. and {Damerdji}, Y. and {Davidson}, M. and {de Teodoro}, P. and {Nu{\~n}ez Campos}, M. and {Delchambre}, L. and {Dell'Oro}, A. and {Esquej}, P. and {Fern{\'a}ndez-Hern{\'a}ndez}, J. and {Fraile}, E. and {Garabato}, D. and {Garc{\'\i}a-Lario}, P. and {Gosset}, E. and {Haigron}, R. and {Halbwachs}, J.-L. and {Hambly}, N.~C. and {Harrison}, D.~L. and {Hern{\'a}ndez}, J. and {Hestroffer}, D. and {Hodgkin}, S.~T. and {Holl}, B. and {Jan{\ss}en}, K. and {Jevardat de Fombelle}, G. and {Jordan}, S. and {Krone-Martins}, A. and {Lanzafame}, A.~C. and {L{\"o}ffler}, W. and {Marchal}, O. and {Marrese}, P.~M. and {Moitinho}, A. and {Muinonen}, K. and {Osborne}, P. and {Pancino}, E. and {Pauwels}, T. and {Recio-Blanco}, A. and {Reyl{\'e}}, C. and {Riello}, M. and {Rimoldini}, L. and {Roegiers}, T. and {Rybizki}, J. and {Sarro}, L.~M. and {Siopis}, C. and {Smith}, M. and {Sozzetti}, A. and {Utrilla}, E. and {van Leeuwen}, M. and {Abbas}, U. and {{\'A}brah{\'a}m}, P. and {Abreu Aramburu}, A. and {Aerts}, C. and {Aguado}, J.~J. and {Ajaj}, M. and {Aldea-Montero}, F. and {Altavilla}, G. and {{\'A}lvarez}, M.~A. and {Alves}, J. and {Anders}, F. and {Anderson}, R.~I. and {Anglada Varela}, E. and {Antoja}, T. and {Baines}, D. and {Baker}, S.~G. and {Balaguer-N{\'u}{\~n}ez}, L. and {Balbinot}, E. and {Balog}, Z. and {Barache}, C. and {Barbato}, D. and {Barros}, M. and {Barstow}, M.~A. and {Bartolom{\'e}}, S. and {Bassilana}, J.-L. and {Bauchet}, N. and {Becciani}, U. and {Bellazzini}, M. and {Berihuete}, A. and {Bernet}, M. and {Bertone}, S. and {Bianchi}, L. and {Binnenfeld}, A. and {Blanco-Cuaresma}, S. and {Blazere}, A. and {Boch}, T. and {Bombrun}, A. and {Bossini}, D. and {Bouquillon}, S. and {Bragaglia}, A. and {Bramante}, L. and {Breedt}, E. and {Bressan}, A. and {Brouillet}, N. and {Brugaletta}, E. and {Bucciarelli}, B. and {Burlacu}, A. and {Butkevich}, A.~G. and {Buzzi}, R. and {Caffau}, E. and {Cancelliere}, R. and {Cantat-Gaudin}, T. and {Carballo}, R. and {Carlucci}, T. and {Carnerero}, M.~I. and {Carrasco}, J.~M. and {Casamiquela}, L. and {Castellani}, M. and {Castro-Ginard}, A. and {Chaoul}, L. and {Charlot}, P. and {Chemin}, L. and {Chiaramida}, V. and {Chiavassa}, A. and {Chornay}, N. and {Comoretto}, G. and {Contursi}, G. and {Cooper}, W.~J. and {Cornez}, T. and {Cowell}, S. and {Crifo}, F. and {Cropper}, M. and {Crosta}, M. and {Crowley}, C. and {Dafonte}, C. and {Dapergolas}, A. and {David}, M. and {David}, P. and {de Laverny}, P. and {De Luise}, F. and {De March}, R.},
        title = "{Gaia Data Release 3. Summary of the content and survey properties}",
      journal = {\aap},
         year = 2023,
        month = jun,
       volume = {674},
          eid = {A1},
        pages = {A1},
          doi = {10.1051/0004-6361/202243940},
archivePrefix = {arXiv},
       eprint = {2208.00211},
 primaryClass = {astro-ph.GA},
       adsurl = {https://ui.adsabs.harvard.edu/abs/2023A&A...674A...1G}
}

@software{Brown_PyGaia_2024,
  author = {Brown, Anthony and others},
  title = {{PyGaia: Python toolkit for Gaia science performance simulation and astrometric catalogue data manipulation.}},
  year = {2024},
  url = {https://github.com/agabrown/PyGaia},
  note = {GitHub repository},
}

@ARTICLE{2023A&A...674A..33G,
       author = {{Gaia Collaboration} and {Montegriffo}, P. and {Bellazzini}, M. and {De Angeli}, F. and {Andrae}, R. and {Barstow}, M.~A. and {Bossini}, D. and {Bragaglia}, A. and {Burgess}, P.~W. and {Cacciari}, C. and {Carrasco}, J.~M. and {Chornay}, N. and {Delchambre}, L. and {Evans}, D.~W. and {Fouesneau}, M. and {Fr{\'e}mat}, Y. and {Garabato}, D. and {Jordi}, C. and {Manteiga}, M. and {Massari}, D. and {Palaversa}, L. and {Pancino}, E. and {Riello}, M. and {Ruz Mieres}, D. and {Sanna}, N. and {Santove{\~n}a}, R. and {Sordo}, R. and {Vallenari}, A. and {Walton}, N.~A. and {Brown}, A.~G.~A. and {Prusti}, T. and {de Bruijne}, J.~H.~J. and {Arenou}, F. and {Babusiaux}, C. and {Biermann}, M. and {Creevey}, O.~L. and {Ducourant}, C. and {Eyer}, L. and {Guerra}, R. and {Hutton}, A. and {Klioner}, S.~A. and {Lammers}, U.~L. and {Lindegren}, L. and {Luri}, X. and {Mignard}, F. and {Panem}, C. and {Pourbaix}, D. and {Randich}, S. and {Sartoretti}, P. and {Soubiran}, C. and {Tanga}, P. and {Bailer-Jones}, C.~A.~L. and {Bastian}, U. and {Drimmel}, R. and {Jansen}, F. and {Katz}, D. and {Lattanzi}, M.~G. and {van Leeuwen}, F. and {Bakker}, J. and {Casta{\~n}eda}, J. and {Fabricius}, C. and {Galluccio}, L. and {Guerrier}, A. and {Heiter}, U. and {Masana}, E. and {Messineo}, R. and {Mowlavi}, N. and {Nicolas}, C. and {Nienartowicz}, K. and {Pailler}, F. and {Panuzzo}, P. and {Riclet}, F. and {Roux}, W. and {Seabroke}, G.~M. and {Th{\'e}venin}, F. and {Gracia-Abril}, G. and {Portell}, J. and {Teyssier}, D. and {Altmann}, M. and {Audard}, M. and {Bellas-Velidis}, I. and {Benson}, K. and {Berthier}, J. and {Blomme}, R. and {Busonero}, D. and {Busso}, G. and {C{\'a}novas}, H. and {Carry}, B. and {Cellino}, A. and {Cheek}, N. and {Clementini}, G. and {Damerdji}, Y. and {Davidson}, M. and {de Teodoro}, P. and {Nu{\~n}ez Campos}, M. and {Dell'Oro}, A. and {Esquej}, P. and {Fern{\'a}ndez-Hern{\'a}ndez}, J. and {Fraile}, E. and {Garc{\'\i}a-Lario}, P. and {Gosset}, E. and {Haigron}, R. and {Halbwachs}, J.-L. and {Hambly}, N.~C. and {Harrison}, D.~L. and {Hern{\'a}ndez}, J. and {Hestroffer}, D. and {Hodgkin}, S.~T. and {Holl}, B. and {Jan{\ss}en}, K. and {Jevardat de Fombelle}, G. and {Jordan}, S. and {Krone-Martins}, A. and {Lanzafame}, A.~C. and {L{\"o}ffler}, W. and {Marchal}, O. and {Marrese}, P.~M. and {Moitinho}, A. and {Muinonen}, K. and {Osborne}, P. and {Pauwels}, T. and {Recio-Blanco}, A. and {Reyl{\'e}}, C. and {Rimoldini}, L. and {Roegiers}, T. and {Rybizki}, J. and {Sarro}, L.~M. and {Siopis}, C. and {Smith}, M. and {Sozzetti}, A. and {Utrilla}, E. and {van Leeuwen}, M. and {Abbas}, U. and {{\'A}brah{\'a}m}, P. and {Abreu Aramburu}, A. and {Aerts}, C. and {Aguado}, J.~J. and {Ajaj}, M. and {Aldea-Montero}, F. and {Altavilla}, G. and {{\'A}lvarez}, M.~A. and {Alves}, J. and {Anderson}, R.~I. and {Anglada Varela}, E. and {Antoja}, T. and {Baines}, D. and {Baker}, S.~G. and {Balaguer-N{\'u}{\~n}ez}, L. and {Balbinot}, E. and {Balog}, Z. and {Barache}, C. and {Barbato}, D. and {Barros}, M. and {Bartolom{\'e}}, S. and {Bassilana}, J.-L. and {Bauchet}, N. and {Becciani}, U. and {Berihuete}, A. and {Bernet}, M. and {Bertone}, S. and {Bianchi}, L. and {Binnenfeld}, A. and {Blanco-Cuaresma}, S. and {Boch}, T. and {Bombrun}, A. and {Bouquillon}, S. and {Bramante}, L. and {Breedt}, E. and {Bressan}, A. and {Brouillet}, N. and {Brugaletta}, E. and {Bucciarelli}, B. and {Burlacu}, A. and {Butkevich}, A.~G. and {Buzzi}, R. and {Caffau}, E. and {Cancelliere}, R. and {Cantat-Gaudin}, T. and {Carballo}, R. and {Carlucci}, T. and {Carnerero}, M.~I. and {Casamiquela}, L. and {Castellani}, M. and {Castro-Ginard}, A. and {Chaoul}, L. and {Charlot}, P. and {Chemin}, L. and {Chiaramida}, V. and {Chiavassa}, A. and {Comoretto}, G. and {Contursi}, G. and {Cooper}, W.~J. and {Cornez}, T. and {Cowell}, S. and {Crifo}, F. and {Cropper}, M. and {Crosta}, M. and {Crowley}, C. and {Dafonte}, C. and {Dapergolas}, A.},
        title = "{Gaia Data Release 3. The Galaxy in your preferred colours: Synthetic photometry from Gaia low-resolution spectra}",
      journal = {\aap},
         year = 2023,
        month = jun,
       volume = {674},
          eid = {A33},
        pages = {A33},
          doi = {10.1051/0004-6361/202243709},
archivePrefix = {arXiv},
       eprint = {2206.06215},
 primaryClass = {astro-ph.SR},
       adsurl = {https://ui.adsabs.harvard.edu/abs/2023A&A...674A..33G}
}

@article{ohman1936red,
  title={The red spectral region of dwarf stars of class M},
  author={{\"O}hman, Yngve},
  journal={Stockholms Observatoriums Annaler, vol. 12, pp. 3.1-3.13},
  volume={12},
  pages={3--1},
  year={1936}
}

@article{thackeray1939intensity,
  title={On the Intensity of Mg 5183 in K-type Stars},
  author={Thackeray, AD},
  journal={Monthly Notices of the Royal Astronomical Society, Vol. 99, p. 492},
  volume={99},
  pages={492},
  year={1939}
}

@ARTICLE{clark1979photoelectric,
       author = {{Clark}, J.~P.~A. and {McClure}, R.~D.},
        title = "{A photoelectric measurement of magnesium for late-type stars.}",
      journal = {\pasp},
         year = 1979,
        month = aug,
       volume = {91},
        pages = {507-518},
          doi = {10.1086/130529},
       adsurl = {https://ui.adsabs.harvard.edu/abs/1979PASP...91..507C}
}

@article{geisler1990washington,
  title={Washington CCD standard fields},
  author={Geisler, Doug},
  journal={Publications of the Astronomical Society of the Pacific},
  volume={102},
  number={649},
  pages={344--350},
  year={1990},
  publisher={The Astronomical Society of the Pacific}
}

@article{majewski2000exploring,
  title={Exploring halo substructure with giant stars. I. Survey description and calibration of the photometric search technique},
  author={Majewski, Steven R and Ostheimer, James C and Kunkel, William E and Patterson, Richard J},
  journal={The Astronomical Journal},
  volume={120},
  number={5},
  pages={2550--2568},
  year={2000}
}

@article{nidever2011discovery,
  title={Discovery of a Large Stellar Periphery Around the Small Magellanic Cloud},
  author={Nidever, David L and Majewski, Steven R and Munoz, Ricardo R and Beaton, Rachael L and Patterson, Richard J and Kunkel, William E},
  journal={The Astrophysical Journal Letters},
  volume={733},
  number={1},
  pages={L10},
  year={2011},
  publisher={The American Astronomical Society}
}

@article{tollerud2012splash,
  title={The SPLASH survey: spectroscopy of 15 M31 dwarf spheroidal satellite galaxies},
  author={Tollerud, Erik J and Beaton, Rachael L and Geha, Marla C and Bullock, James S and Guhathakurta, Puragra and Kalirai, Jason S and Majewski, Steve R and Kirby, Evan N and Gilbert, Karoline M and Yniguez, Basilio and others},
  journal={The Astrophysical Journal},
  volume={752},
  number={1},
  pages={45},
  year={2012},
  publisher={The American Astronomical Society}
}

@article{gilbert2012global,
  title={Global Properties of M31's Stellar Halo from the SPLASH Survey. I. Surface Brightness Profile},
  author={Gilbert, Karoline M and Guhathakurta, Puragra and Beaton, Rachael L and Bullock, James and Geha, Marla C and Kalirai, Jason S and Kirby, Evan N and Majewski, Steven R and Ostheimer, James C and Patterson, Richard J and others},
  journal={The Astrophysical Journal},
  volume={760},
  number={1},
  pages={76},
  year={2012},
  publisher={The American Astronomical Society}
}

@article{majewski2008discovery,
  title={Discovery of an extended, halo-like stellar population around the Large Magellanic Cloud},
  author={Majewski, Steven R and Nidever, David L and Munoz, Ricardo R and Patterson, Richard J and Kunkel, William E and Carlin, Jeffrey L},
  journal={Proceedings of the International Astronomical Union},
  volume={4},
  number={S256},
  pages={51--56},
  year={2008},
  publisher={Cambridge University Press}
}

@article{zasowski2013target,
  title={Target selection for the apache point observatory Galactic evolution experiment (APOGEE)},
  author={Zasowski, G and Johnson, Jennifer A and Frinchaboy, PM and Majewski, Steven Raymond and Nidever, DL and Pinto, HJ Rocha and Girardi, L and Andrews, B and Chojnowski, Stephen Drew and Cudworth, KM and others},
  journal={The Astronomical Journal},
  volume={146},
  number={4},
  pages={81},
  year={2013},
  publisher={The American Astronomical Society}
}

@article{beaton2021final,
  title={Final Targeting Strategy for the Sloan Digital Sky Survey IV Apache Point Observatory Galactic Evolution Experiment 2 North Survey},
  author={Beaton, Rachael L and Oelkers, Ryan J and Hayes, Christian R and Covey, Kevin R and Chojnowski, SD and De Lee, Nathan and Sobeck, Jennifer S and Majewski, Steven R and Cohen, Roger E and Fernandez-Trincado, Jose and others},
  journal={The Astronomical Journal},
  volume={162},
  number={6},
  pages={302},
  year={2021},
  publisher={The American Astronomical Society}
}

@article{morrison2001mapping,
  title={Mapping the galactic halo. IV. Finding distant giants reliably with the Washington system},
  author={Morrison, Heather L and Olszewski, Edward W and Mateo, Mario and Norris, John E and Harding, Paul and Dohm-Palmer, Robbie C and Freeman, Kenneth C},
  journal={The Astronomical Journal},
  volume={121},
  number={1},
  pages={283--294},
  year={2001}
}

@article{casey2018infrared,
  title={Infrared colours and inferred masses of metal-poor giant stars in the Kepler field},
  author={Casey, Andrew R and Kennedy, Grant M and Hartle, Tom R and Schlaufman, Kevin C},
  journal={Monthly Notices of the Royal Astronomical Society},
  volume={478},
  number={2},
  pages={2812--2818},
  year={2018},
  publisher={Oxford University Press}
}

@PHDTHESIS{2007PhDT.........7T,
       author = {{Teig}, Matthew James},
        title = "{Developing a technique to separate dwarfs and giants using DDO51 photometry in a photometric survey of M33}",
       school = {University of California, Irvine},
         year = 2007,
        month = jan,
       adsurl = {https://ui.adsabs.harvard.edu/abs/2007PhDT.........7T}
}
\bibliographystyle{aasjournalv7}



\end{document}